\documentclass[twocolumn]{aastex7}
\usepackage{apjfonts}
\usepackage{nccmath}
\usepackage{stackengine}
\usepackage{natbib}
\usepackage[htt]{hyphenat}
\usepackage{blindtext}
\usepackage{rotating}
\usepackage{graphicx}
\usepackage{graphics,graphicx,float,import,enumitem,booktabs}

\newcommand{\asymuncert}[3]{\ensuremath{#1^{+#2}_{-#3}}}

\newcommand\te{T$_e$}

\shorttitle{CECILIA: The Mass-Metallicity Relation of Low-Mass Galaxies at Cosmic Noon}

\shortauthors{Raptis et al.}

\begin{document}

\title{CECILIA: The Mass-Metallicity Relation of Low-Mass Galaxies at Cosmic Noon}

\author[0009-0008-2226-5241]{Menelaos Raptis}
\affiliation{Department of Physics and Astronomy, Franklin \& Marshall College, 637 College Avenue, Lancaster, PA 17603, USA}
\affiliation{The Observatories of the Carnegie Institution for Science, 813 Santa Barbara Street, Pasadena, CA 91101, USA}
\email{mraptis@fandm.edu}

\author[0000-0002-8459-5413]{Gwen C. Rudie}
\affiliation{The Observatories of the Carnegie Institution for Science, 813 Santa Barbara Street, Pasadena, CA 91101, USA}
\email{gwen@carnegiescience.edu}

\author[0000-0002-6967-7322]{Ryan F. Trainor}
\affiliation{Department of Physics and Astronomy, Franklin \& Marshall College, 637 College Avenue, Lancaster, PA 17603, USA}
\affiliation{William H. Miller III Department of Physics and Astronomy, Johns Hopkins University, Baltimore, MD 21218, USA}
\email{ryan.trainor@fandm.edu}

\author[0000-0002-0361-8223]{Noah S. J. Rogers}
\affiliation{Center for Interdisciplinary Exploration and Research in Astrophysics (CIERA), Northwestern University, 1800 Sherman Ave., Evanston, IL 60201, USA}
\email{noah.rogers@northwestern.edu}

\author[0000-0001-6369-1636]{Allison L. Strom}
\affiliation{Center for Interdisciplinary Exploration and Research in Astrophysics (CIERA), Northwestern University, 1800 Sherman Ave., Evanston, IL 60201, USA}
\affiliation{Department of Physics and Astronomy, Northwestern University, 2145 Sheridan Road, Evanston, IL 60208, USA}
\email{allison.strom@northwestern.edu}

\author[0000-0003-2385-9240]{Nathalie A. Korhonen Cuestas}
\affiliation{Center for Interdisciplinary Exploration and Research in Astrophysics (CIERA), Northwestern University, 1800 Sherman Ave., Evanston, IL 60201, USA}
\affiliation{Department of Physics and Astronomy, Northwestern University, 2145 Sheridan Road, Evanston, IL 60208, USA}
\email{NathalieKorhonenCuestas2029@u.northwestern.edu}

\author[0000-0002-6034-082X]{Caroline von Raesfeld}
\affiliation{Center for Interdisciplinary Exploration and Research in Astrophysics (CIERA), Northwestern University, 1800 Sherman Ave., Evanston, IL 60201, USA}
\affiliation{Department of Physics and Astronomy, Northwestern University, 2145 Sheridan Road, Evanston, IL 60208, USA}
\email{CarolinevonRaesfeld2027@u.northwestern.edu}

\author[0009-0009-0946-0605]{Ye Lin}
\affiliation{Department of Physics and Astronomy, Franklin \& Marshall College, 637 College Avenue, Lancaster, PA 17603, USA}
\affiliation{Department of Physics and Astronomy, University of California, Riverside, CA 92521, USA}
\email{ylin375@ucr.edu}

\author[0009-0005-8667-4771]{Ojima Ojodomo Abraham}
\affiliation{Department of Physics and Astronomy, Franklin \& Marshall College, 637 College Avenue, Lancaster, PA 17603, USA}
\email{abrahamojima0@gmail.com}

\author[0009-0007-5328-4720]{Christopher Chapman}
\affiliation{Department of Physics and Astronomy, Franklin \& Marshall College, 637 College Avenue, Lancaster, PA 17603, USA}
\email{cwmchapman@gmail.com}

\author[0000-0002-4834-7260]{Charles C. Steidel}
\affiliation{Cahill Center for Astronomy and Astrophysics, California Institute of Technology, MS 249-17, Pasadena, CA 91125, USA}
\email{ccs@astro.caltech.edu}

\author[0000-0003-0695-4414]{Michael V. Maseda}
\affiliation{Department of Astronomy, University of Wisconsin-Madison, 475 N. Charter Street, Madison, WI 53706, USA}
\email{maseda@astro.wisc.edu}

\begin{abstract}
A galaxy's metallicity and its relation to stellar mass encode the history of gas accretion, star formation, and outflows within cosmic ecosystems. We present new constraints on the low-mass end of the mass–metallicity relation (MZR) at $z\sim2-3$ from ultra-deep JWST/NIRSpec spectroscopy of seven continuum-faint galaxies in the Chemical Evolution Constrained using Ionized Lines in Interstellar Aurorae (CECILIA) Faint sample \citep{Raptis2025}. Our sample includes Ly$\alpha$-selected and other low-luminosity star-forming galaxies with low stellar masses $\log(M_\ast/M_\odot) \approx 7.2$--$9.7$ and moderately faint rest-UV magnitudes ($-20.7 \lesssim M_{\rm UV} \lesssim -17.3$). Gas-phase oxygen abundances, calculated using empirical calibrations of [O \textsc{iii}]/H$\beta$ together with [N\,\textsc{ii}]/H$\alpha$ constraints, span $\sim0.04$--$0.5\,Z_\odot$. We measure a steep MZR slope of $\gamma = 0.48 \pm 0.11$, suggesting a rapid increase in metal retention efficiency with mass, consistent with energy-driven outflows. Comparison with lower- and higher-redshift studies indicates an evolution in normalization from $z \sim 0$ to $z \sim 2$, reflecting less metal enrichment in early galaxies. We find no significant evolution in the MZR between $z\sim2$ and the Epoch of Reionization, suggesting that our galaxies may serve as useful analogs of reionization-era systems. Expanded samples and direct $T_e$-based abundance measurements will be crucial to fully trace the build-up of metals in low-mass galaxies during the peak epoch of cosmic star formation and to test the reliability of strong-line calibrations in these galaxies.
\end{abstract}


\section{Introduction}
\label{Introduction}
 
Metals, produced as byproducts of stellar nucleosynthesis, accumulate over time through successive generations of star formation. Their distribution is further influenced by a galaxy's interaction with its environment via gas inflows, feedback-driven outflows, and the recycling of enriched material. The observed gas-phase metallicity thus reflects the net balance between metal production, dilution, and redistribution. This observed gas-phase metallicity, in turn, offers insights into the galaxy's star formation history, baryon cycle, and broader context of cosmic chemical evolution.

The mass-metallicity relation (MZR) — the observed correlation between a galaxy’s stellar mass and gas-phase oxygen abundance — is a key diagnostic of chemical evolution, as galaxies build up their stellar mass while enriching their interstellar medium with heavy elements. First identified in local galaxies by \citet{Lequeux1979}, who found a positive correlation between the total mass of the galaxy and the oxygen abundance, the MZR has since become a cornerstone of galaxy evolution studies \citep[see][for a review]{2019A&ARv..27....3M}.


In the local Universe, the MZR is well characterized, particularly for massive star-forming galaxies. \citet{trem2004} used 53,000 star-forming galaxies from the Sloan Digital Sky Survey Early Data Release (SDSS; \citealt{York2000, Abazajian2003}), to establish a strong, monotonic increase in gas-phase oxygen abundance with stellar mass, which flattens at the high-mass end around $\log(M_*/M_\odot) \sim 10.5$. They also found a relatively small intrinsic scatter of $\sim$0.1~dex in metallicity at fixed stellar mass, a result that has since been confirmed by numerous follow-up studies \citep[e.g.,][]{Mannucci2010, Bezanson2015, Curti2020}.

Gas-phase metallicity is estimated using a variety of calibrations and emission-line ratios, which can result in systematic offsets between different analyses of the MZR. Recognizing this, subsequent studies explored how the choice of metallicity diagnostic impacts the shape and normalization of the MZR. \citet{Kewley2008} compared several strong-line methods — using bright nebular lines such as [\ion{O}{2}]~$\lambda\lambda3727,3729$, H$\beta$, [\ion{O}{3}]~$\lambda5007$, [\ion{N}{2}]~$\lambda6585$, and H$\alpha$ — and showed that absolute metallicities can vary by up to 0.7 dex depending on the calibration used, although the relative trends remain consistent. \citet{Andrews2013} employed direct electron temperature (\te)-based metallicities using stacked SDSS spectra, providing a complementary calibration of the MZR at $z \sim 0$. Their work confirmed a similar slope but found a systematically lower normalization compared to strong-line estimates, reflecting the known offset between \te-based abundances and photoionization-model strong-line calibrations (e.g., \citealt{mous2010}).

\citet{Erb2006} established the existence of a mass-metallicity relation at $z \sim 2$, demonstrating that galaxies at this epoch have systematically lower metallicities than local galaxies of similar stellar mass. More recent surveys, particularly those employing advanced ground-based near-infrared spectrographs such as Keck/MOSFIRE, VLT/X-Shooter, and VLT/KMOS have provided deeper insights into galaxy populations during the epoch of peak star formation (2~$\lesssim z \lesssim$~3, \citealt{Madau2014}), often referred to as Cosmic Noon \citep[e.g.,][]{stei2014, 2014A&A...563A..58T, 2015ApJ...799..209W, Shapley2019, Sanders2021, Henry2021, Strom2022}. These instruments, while highly effective, face limitations inherent to ground-based observations — such as atmospheric transmission and sky background — which make it challenging to detect faint galaxies, limiting observations to more massive and luminous systems.

\begin{deluxetable*}{lccccc}
\tablecaption{Galaxy Names and Properties \label{t:galaxies}}
\tablewidth{\textwidth}
\tablehead{
  \colhead{Galaxy Name} &
  \colhead{Redshift} &
  \colhead{$\log(M_*/M_\odot)$}&
  \colhead{M$_{\text{UV}}$} &
  \colhead{SFR (M$_{\odot}$/year)} &
  \colhead{12 + log(O/H)}
}
\startdata
Q2343-NB2929 & 2.546 & $8.89^{+0.16}_{-0.22}$ & $-18.99$ & \asymuncert{1.27}{0.85}{0.34} & $7.93 \pm 0.21$ \\
Q2343-NB2089 & 2.571 & $7.69^{+0.45}_{-0.45}$ & $< -17.29$ & \asymuncert{1.54}{0.49}{0.29} & $7.30 \pm 0.06$ \\
Q2343-NB2875 & 2.545 & $7.24^{+0.53}_{-0.34}$ & $< -17.74$ & \asymuncert{0.86}{0.15}{0.11} & $7.36 \pm 0.05$ \\
Q2343-NB2571 & 2.577 & $8.52^{+0.44}_{-0.29}$ & $-17.21$ & \asymuncert{0.64}{0.14}{0.09} & $7.93 \pm 0.17$ \\
Q2343-fBM47  & 2.278 & $9.23^{+0.12}_{-0.15}$ & $-20.65$ & \asymuncert{3.04}{1.33}{0.67} & $8.13 \pm 0.10$ \\
Q2343-fBM40  & 2.147 & $9.18^{+0.16}_{-0.18}$ & $-20.43$ & \asymuncert{3.65}{1.23}{0.70} & $8.26 \pm 0.08$ \\
Q2343-fC23   & 2.173 & $9.42^{+0.09}_{-0.13}$ & $-19.85$ & \asymuncert{4.51}{5.89}{1.45} & $8.36 \pm 0.13$ \\
\enddata

\tablecomments{Names and properties of the galaxies included in this investigation. Stellar masses assume a \citet{Kroupa2001}-like IMF. SFRs are derived from dust-corrected H$\alpha$ luminosities using metallicity-dependent conversion factors from \citet{korh2025}, as described in Section~\ref{sec:m_sfr}. The metallicities listed in the final column are calculated using the O3-based calibration curve from \citet{Nakajima22} as described in Section~\ref{sec:gas_phase}.}
\label{table:properties}
\end{deluxetable*}


Low-mass ($M_\star < 10^{8.5}\,M_\odot$), faint galaxies are particularly valuable for testing the limits of the MZR because their shallow gravitational potentials make them more susceptible to metal loss via galactic winds \citep[e.g.,][]{Dekel1986, Mac_Low1999}. As a result, deviations from the typical MZR (i.e., the relation followed by more massive galaxies at similar redshift) at the low-mass end may manifest as both systematically lower metallicities --- which would steepen the MZR slope --- and increased scatter at fixed stellar mass. Studying these systems helps determine whether the physical processes that govern chemical evolution operate consistently across the full mass range or if different mechanisms dominate in the low-mass regime.


Studies of the MZR in local, low-mass galaxies have generally shown that the MZR extends to stellar masses as low as $\log(M_\star/M_\odot) \sim 6$--$8$ with relatively small intrinsic scatter comparable to that observed at higher masses \citep[e.g.,][for $z \sim 0$]{Lee2006}. Stellar metallicity-based measurements have likewise extended the MZR to very low-mass dwarf galaxies in the Local Group, with \citet{Kirby2013} finding a tight relation down to $\log(M_\star/M_\odot) \sim 3$. Observations across a broad redshift range reveal a strong evolution of the MZR over cosmic time, with metallicities in low-mass galaxies evolving more rapidly than in their massive counterparts, supporting a “chemical downsizing” scenario where massive galaxies enrich their interstellar medium early, while low-mass galaxies build up metals over more extended timescales \citep[e.g.,][]{Maiolino2008, Henry2013}.

Star formation efficiency tends to increase with stellar mass and has been argued to be a dominant driver of the MZR, as low-mass galaxies are generally more gas-rich and less efficient at converting gas into stars and metals \citep{Hughes2013}. While galactic winds and metal loss may contribute to shaping the MZR, several studies suggest that these mechanisms alone cannot fully explain the relatively tight relation and low scatter observed at the low-mass end \citep[e.g.,][]{Lee2006, Hughes2013}.



The unprecedented infrared sensitivity of JWST has enabled robust measurements of the MZR to be extended to earlier cosmic epochs and lower stellar masses, allowing recent studies to probe low-mass galaxies at $z > 2$ that were inaccessible with earlier facilities \citep[e.g.,][]{naka2023, Li2023, Scholte2025}.


In this paper, we investigate the MZR of low-mass, continuum-faint star-forming galaxies at $z \sim 2$--3. Specifically, we focus on seven of the faintest galaxies in the CECILIA sample \citep{stro2023, Raptis2025}, including Ly$\alpha$-selected and other continuum-faint sources that have comparatively low stellar masses with an average of $\log(M_\star/M_\odot) \sim 8.5$. The companion study, \citet{Raptis2025}, examines the rest-optical emission line ratios and excitation conditions of this same CECILIA subsample. Unlike previous JWST studies, which often rely on stacked spectra or focus on more luminous systems, our analysis uses ultra-deep NIRSpec/MSA spectroscopy ($29.5$ hr in G235M/F170LP) to probe individual galaxies at low mass and low luminosity, enabling a detailed exploration of the ionized gas and chemical enrichment in galaxies that have so far been underrepresented in MZR studies.


This manuscript is organized as follows: In \S2, we describe the JWST observations, data, and sample selection. In \S3, we present the stellar masses, star formation rates, and gas-phase metallicities, including the calibrations used in this paper. In \S4, we compare the slope of our fits in the MZR plane with those from other studies and discuss the redshift evolution in the MZR. Finally, we conclude in \S5.

In this work, we assume a $\Lambda$CDM cosmology with $H_0$$=$70 km s$^{-1}$ Mpc$^{-1}$, $\Omega_m$$=$0.3, and $\Omega_\Lambda$$=$0.7. We adopt the solar abundance values from \citet{aspl2021}: 12+log(O/H)$_\odot = 8.69\pm0.04$ dex. Throughout the text, we refer to emission lines using their vacuum wavelengths and we use the term "metallicity" to refer to the gas-phase oxygen abundance.






\section{Data}
\label{Data}

\subsection{Spectroscopic measurements}
\label{sec:spectra}

The galaxies examined in this work are part of the CECILIA program, a JWST/NIRSpec multi-object spectroscopic survey focused on star-forming galaxies at $z \sim 2$--3 within the Q2343 field \citep{stro2023}. A central objective of CECILIA is to detect multiple faint rest-optical auroral emission lines, enabling detailed studies of the ionized gas conditions and chemical enrichment in high-redshift galaxies. These observations span the rest-optical to near-infrared wavelengths ($\lambda_{\text{rest}} \approx 0.5$--$1.2\,\mu$m), covering many key diagnostics at Cosmic Noon.

A full description of the sample selection, observational strategy, and data reduction procedures is presented in \citet{stro2023}, \citet{Rogers2024}, and \cite{Rogers2025}. In brief, targets were drawn from the Keck Baryonic Structure Survey (KBSS; \citealt{rudi2012, stei2014, stro2017}) and KBSS-Ly$\alpha$ \citep{trai2015, Trainor2016} surveys, with selection based on available rest-UV and rest-optical spectra as well as photoionization-model predictions of auroral line strengths. Observations were obtained with JWST/NIRSpec using the G235M/F170LP ($29.5$ hr) and G395M/F290LP ($1.1$ hr) grating configurations.

This paper focuses on a subsample of seven continuum-faint galaxies drawn from the full CECILIA sample. The subsample includes four narrowband (NB)-selected Ly$\alpha$ emitters (LAEs) and three faint UV-selected Lyman break galaxies (fLBGs). Further details on the sample definition and individual galaxy properties can be found in the companion study \citep{Raptis2025}. The sample spans a redshift range of $z = 2.1$--2.6, with rest-frame UV magnitudes of $M_{\mathrm{UV}} \sim -19$ and H$\alpha$-based star formation rates of $\sim$0.6--4.5~$M_\odot$~yr$^{-1}$, derived from the H$\alpha$ emission line luminosity, as described in Section~\ref{sec:m_sfr}.

Spectral extraction and emission line fitting of the JWST/NIRSpec spectroscopy follow the methodology detailed in \cite{Raptis2025}. The 1D spectra are optimally extracted following the procedure of \citet{Horne1986}, and emission line fluxes are measured by fitting a composite model consisting of Gaussian emission lines and a polynomial continuum. Flux uncertainties are estimated using a Markov Chain Monte Carlo (MCMC) approach. The emission lines typically detected at signal-to-noise ratios (S/N) greater than 2 in most of our galaxies include H$\beta$, [O~\textsc{iii}]~$\lambda4960$, [O~\textsc{iii}]~$\lambda5007$, He~\textsc{i}~$\lambda5877$, H$\alpha$, [S~\textsc{ii}]~$\lambda\lambda6718,6733$, He~\textsc{i}~$\lambda7067$, and [Ar~\textsc{iii}]~$\lambda7138$. These galaxies generally do not exhibit significant [N~\textsc{ii}]~$\lambda6585$ emission, a signature of their low gas-phase metallicities and high-ionization conditions within the interstellar medium (see \citealt{Raptis2025}).

\subsection{Photometry}
\label{sec:photometry}

As part of KBSS, the CECILIA JWST/NIRSpec \citep{Jakobsen2022} field has substantial ancillary imaging, including Keck/LRIS \citep{oke1995,steidel2004}, HST/WFC3, and Spitzer/IRAC measurements. For this study, we use photometry from the LRIS $B$, $G$, and $\mathcal{R}_s$ filters, the WFC3 F110W, F140W, and F160W\footnote{The F110W and F160W images of the CECILIA field have incomplete spatial coverage, so not all the galaxies in our sample have photometry in these bands.} filters, and the IRAC Ch2 ($4.5\,\mu$m) filter. The LRIS NB4325 narrowband filter was also used for the initial selection of the NB sources in this study as described by \citet{Trainor2025}, although it is not included in any photometric analysis for this study.

For all bands except the IRAC Ch2 image, each image was smoothed to match the seeing in the LRIS $\mathcal{R}_s$ band, and then Source Extractor \citep{bertin1996} was used in two-image mode to extract the isophotal magnitudes in each band, using the smoothed F140W image for source identification. The color difference between the Source Extractor isophotal and `auto' magnitudes in the F140W band was used to correct for light beyond the isophotal aperture. Objects NB2089, NB2571, and NB2875 are not well-detected in the F140W image, so for these sources we used the original Ly$\alpha$ narrowband detection aperture to define the extraction apertures for the broadband photometry.  Flux uncertainties were determined by simulations of injected sources with a range of sizes, which results in an uncertainty estimate per image for typical source sizes. This per-image flux uncertainty was then scaled according to the local noise and isophotal aperture size for each source.

The IRAC2 photometry was performed by fitting a multi-component point spread function (PSF)-convolved model to each target source and other F140W-detected sources in the surrounding $\sim$10$''$ region. Flux uncertainties were calculated by repeatedly re-fitting the object fluxes after adding a spatially shifted map of the residuals to the model fit to account for the sampling-dependent variations in the IRAC point response function.

\begin{figure*}[t]
   \centering
   \includegraphics[width=1\textwidth]{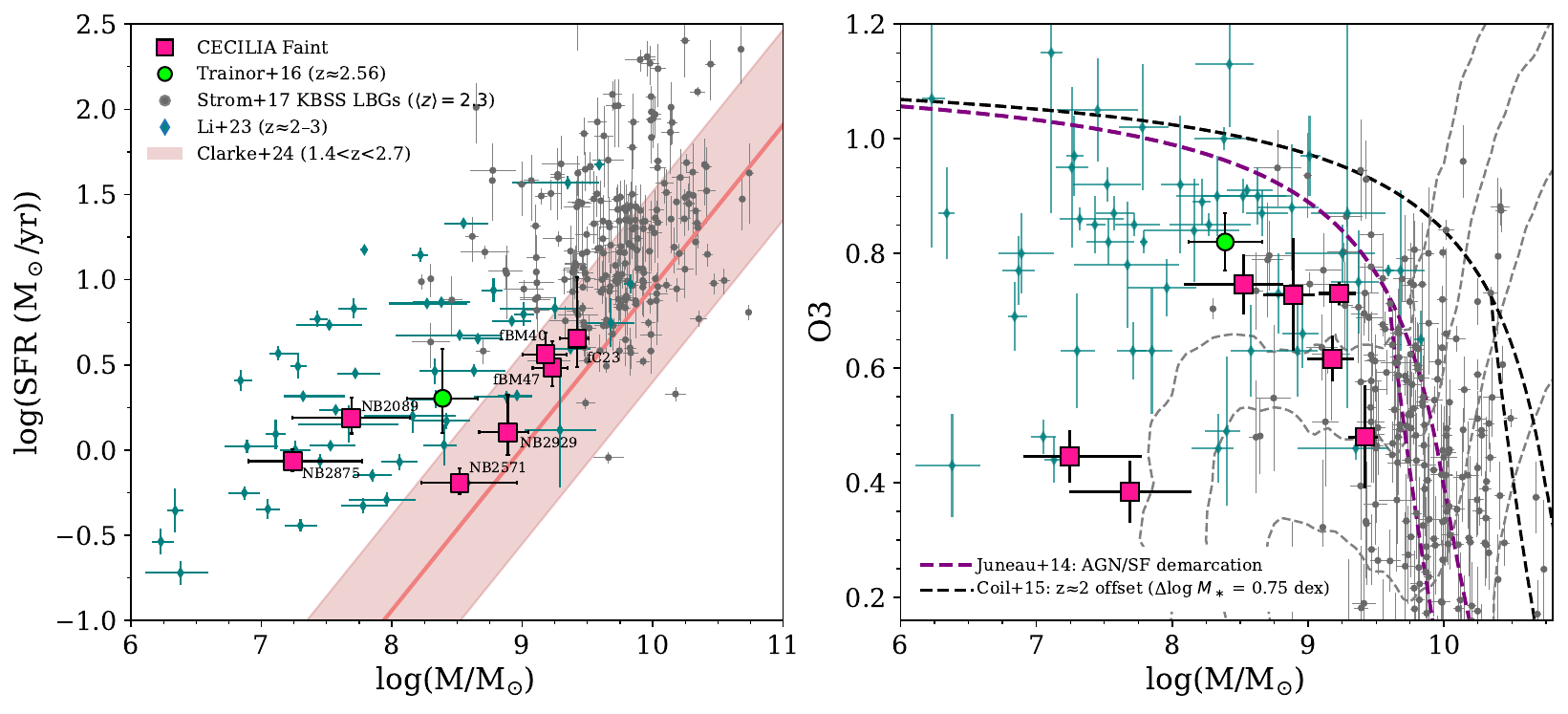}
   \caption{\textit{Left:} H$\alpha$-based SFRs versus stellar mass for our CECILIA Faint sample (pink squares), derived using a metallicity-dependent conversion. For consistency, we recalculate the SFRs of the $z \sim 2.56$ stack from \citet[][green circle]{Trainor2016} and of the KBSS LBGs (\citealt{stro2017}; gray circles) using the same method. For comparison, we also show the H$\alpha$-based SFR--$M_\star$ relation for $1.4 < z < 2.7$ galaxies from \citet{Clarke2024}, as well as  individual lensed dwarf galaxies at $z \sim 2$--3 selected as [O\,\textsc{iii}] emitters in JWST/NIRISS spectroscopy by \citet[][teal diamonds]{Li2023}. We generally find consistency between the CECILIA Faint sources and these comparison samples in the SFR-M$_*$ plane. \textit{Right:} The MEx diagram for the CECILIA Faint galaxies and the same comparison samples as in the left panel. The gray dashed contours represent SDSS $z \sim 0$ galaxies. We include the original \citet{Juneau2014} star-forming/AGN boundaries (purple dashed) and the redshift-adjusted versions shifted by $\Delta\log M_\ast = 0.75$~dex from \citet{Coil2015}, appropriate for $z \sim 2$ galaxies. The CECILIA Faint galaxies broadly agree with the \citet{Li2023} sample at $M_\ast > 10^{8}\,M_\odot$, but have significantly lower O3 ratios at lower masses.}

   \label{fig:MS}
\end{figure*}

The above steps were performed for all the CECILIA Faint sources in Table~\ref{table:properties}, as well as for the other sources in the Q2343 field that form the parent sample for the NB objects described in this paper to make a raw flux catalog. Using the resulting photometric catalog, we then construct a catalog of emission-line-corrected fluxes for each source using the JWST/NIRSpec spectra as well as previous MOSFIRE measurements where present, including the stacked MOSFIRE $H$ and $K$ spectra from the KBSS-Ly$\alpha$ parent sample \citep{Trainor2016} and a new $J$-band stack (Y. Lin et al., in prep.) that includes coverage of [\ion{O}{2}]$\lambda\lambda$3727,29.  We correct the photometry for contamination from strong lines in the range $3700\,\textrm{\AA}\lesssim \lambda_\textrm{rest}\lesssim 2\,\mu\textrm{m}$, including the [\ion{O}{2}], [\ion{O}{3}], [\ion{N}{2}], [\ion{S}{2}], [\ion{S}{3}], Balmer, and Paschen series emission lines in this range. To estimate the correction factor for each line and source, we use the observed JWST/NIRSpec equivalent width, the observed MOSFIRE flux, or the stacked MOSFIRE flux scaled to the source's H$\alpha$ flux in decreasing order of preference. For the lines at $\lambda_\mathrm{rest} > 9000~\text{\AA}$, we assume the measured line ratios relative to H$\alpha$ from \citet{Rogers2024}.

Photometric uncertainties in the line-corrected catalog are calculated by propagating the line flux uncertainties for the measured or estimated line fluxes. For line fluxes estimated from a scaled stack spectrum, the standard deviation of flux ratios relative to H$\alpha$ among objects with a direct measurement for that line is used to estimate line-flux uncertainty. The slit-loss correction uncertainty (which is $\sim$30\% for some of the MOSFIRE measurements) is also included.

Finally, in order to compare our measurements with typical properties of the KBSS-Ly$\alpha$ parent sample from \citet{Trainor2016}, we perform the same process on stacked images of all the KBSS-Ly$\alpha$ objects that have HST/WFC3-F140W coverage\footnote{Note that the photometric stack does not include the 8 objects in \citet{Trainor2016} that are in the Q1603 field.}, extracting a stacked flux in each band and performing emission-line corrections as above. We discuss our inferences from the emission-line corrected photometry for the CECILIA Faint objects and the KBSS-Ly$\alpha$ stack in Section~\ref{sec:m_sfr}.

\section{Analysis and Results}
\label{Analysis}

\subsection{Stellar Mass and Star Formation Rate}
\label{sec:m_sfr}

Stellar masses were determined by fitting spectral-energy distribution (SED) models to the emission-line corrected photometry described above. We fit the SED models with BAGPIPES \citep{carnall2018}, using the BPASS v2.2.1 \citep{Stanway2018} stellar population models. We assume a \citet{Kroupa2001}-like initial mass function (IMF) with an upper mass limit of 100$\,$M$_\odot$, with nebular continuum contribution from Cloudy \citep{ferl2013}. Because the photometry has already been corrected using our empirical emission line measurements, we omit the predicted emission-line contributions to our SED models during fitting. We assume an ionization parameter $\log(U)=-2.5$ and a flat metallicity prior in  $[0,\,0.5]\,\mathrm{Z}_\odot$ based on the measurements of \citet{Trainor2016}, although in practice neither the ionization parameter nor metallicity appreciably affect our SED fits, given that the emission line fluxes are modeled empirically.

We adopt a continuity star-formation model with star-formation history (SFH) bins defined by (0, 10 Myr, 30 Myr, 100 Myr, 300 Myr, 1 Gyr, 2 Gyr, 3 Gyr, 4 Gyr), with the highest-age bins omitted when they exceed the age of the Universe at the spectroscopically-measured redshift. Rather than adopting a prior on the total mass formed over the integrated SFH, we modify the Bagpipes SFH module to impose a prior only on the most recent SFH bin $(0,\,10~\mathrm{Myr})$, for which we set a Gaussian prior based on the dust-corrected H$\alpha$-based star formation rate (SFR) and its uncertainty for each galaxy as reported in Table \ref{table:properties}. The SFRs are derived from dust-corrected H$\alpha$ luminosities following the approach of \citet{Raptis2025}, but using metallicity-dependent conversion factors from \citet{korh2025}, where the stellar metallicity matches the individual gas-phase metallicities measured in Section~\ref{sec:gas_phase}. In practice, this prior produces a lower bound on the stellar masses of order $M_*\gtrsim (10\,\textrm{Myr})\times\textrm{SFR}$. For the majority of our sources, adopting this prior produces a mass posterior very similar to a continuity SFH with no SFR prior (i.e., substituting a flat mass prior over $\log(M_*/\textrm{M}_\odot)$ in $[0,13]$ yields a similar mass). However, the two faintest NB sources (NB2089 and NB2875) have weak detections or non-detections in the emission-line-corrected broadband photometry, such that they are formally consistent with arbitrarily small masses in the absence of realistic constraints from the SFRs or emission line measurements.

Finally, we adopt an SMC \citep{Gordon2003} dust law as in \citet{korh2025}. For most of our sources we set only a flat prior on $A_V$ in $[0,\,1]\,$mag, but for our sources with Ly$\alpha$ detections, we set an upper limit on $A_V$ based on the requirement that the dust model should not predict a lower differential attenuation of Ly$\alpha$ with respect to H$\alpha$ than the observed Ly$\alpha$/H$\alpha$ ratio under the assumption of Case-B recombination. We regard this as a conservative limit, given that Ly$\alpha$ is sensitive to other sources of attenuation and scattering (i.e., \ion{H}{1}) in addition to dust. In practice, this constraint prevents Bagpipes from preferring massive, dust-obscured model galaxies that are technically allowed by our weak broadband limits, but which are inconsistent with the observed  Ly$\alpha$ emission. We return to the effects of our SFR and $A_V$ priors in Section \ref{sec:slope_at_cosmic_noon}. In addition to fitting the above model to each of the CECILIA Faint sources, we perform the same fitting procedure to photometric measurements of the stacked parent sample of KBSS-Ly$\alpha$ sources described in Section~\ref{sec:photometry}. Each stellar mass reported in Table~\ref{table:properties} represents the median of the Bagpipes posterior distribution for that source, with uncertainties given by the 16th and 84th percentiles. To assess the reliability of our stellar mass estimates, we derived independent dynamical masses using galaxy sizes and velocity dispersions and find good agreement between the two sets of estimates. Additional details are presented in Appendix~\ref{appendix:dynamical_masses}.



In the left panel of Figure~\ref{fig:MS}, we present the star-forming main sequence for our CECILIA Faint galaxies, shown as pink squares. We also include several comparisons at similar redshifts, including the KBSS-Ly$\alpha$ parent-sample stack (green circle; \citealt{Trainor2016}), the KBSS LBGs (gray circles; \citealt{korh2025}), the \citet{Clarke2024} SFMS relation (orange shaded band), and low-mass, line-selected lensed galaxies from \citet{Li2023} (teal diamonds). For our sample, the KBSS-Ly$\alpha$ stack, and the KBSS LBGs, the H$\alpha$-based SFRs are derived from the H$\alpha$ luminosity using
$\log_{10}[\mathrm{SFR},(\mathrm{M_\odot,yr^{-1}})] = \log_{10}[L_{\mathrm{H}\alpha},(\mathrm{erg,s^{-1}})] + C$, where $C$ is a metallicity-dependent conversion factor following the methodology of \citet{korh2025}, who assumed an [O/Fe] offset of 0.35 in order to relate the nebular O/H measurements to the Fe/H-dependent stellar models. For our sources, with metallicities determined in Section~\ref{sec:gas_phase}, $C$ lies in the range $(-41.68, -41.62)$. The KBSS LBG stellar masses used in this work are the same as in \citet{korh2025}; like our CECILIA Faint masses, they are obtained with \textsc{BAGPIPES} using BPASS binary stellar population models, an SMC dust law, and a continuity SFH\footnote{As discussed in \citet{korh2025}, these assumptions yield stellar masses $\sim$0.06\,dex higher on average than the constant-SFH masses of \citet{Theios2019}, although individual galaxies may have higher or lower masses depending on the shape of the SFH.}. In \citet{Clarke2024}, stellar masses were derived assuming a delayed-$\tau$ SFH, while SFRs were obtained using an H$\alpha$-based, metallicity-dependent BPASS calibration, which corresponds to a $C$ range of approximately $(-41.59, -41.37)$ --- slightly higher than the range adopted for our CECILIA Faint galaxies. The lensed dwarf galaxies from \citet{Li2023} were identified with JWST/NIRISS slitless spectroscopy on the basis of significant detection of their [O\,\textsc{iii}]~$\lambda5007$ emission (S/N $> 10$); their stellar masses were derived assuming a constant SFH, and their SFRs were calculated using the solar metallicity \citet{Kennicutt1998} relation with a \citet{Chabrier2003} IMF, corresponding to $C = -41.34$. This value is on average 0.32 dex higher than the $C$ values we adopt for our CECILIA Faint galaxies. Most of our CECILIA Faint galaxies fall within the scatter of the \citet{Clarke2024} SFMS relation, providing a useful point of comparison across different studies, while several KBSS galaxies and many of the \citet{Li2023} low-mass systems lie above this relation. Notably, our two lowest-mass galaxies also fall above the Clarke relation but remain consistent with many of the \citet{Li2023} dwarf galaxies.

\begin{figure*}[t]
   \centering
   \includegraphics[width=1\textwidth]{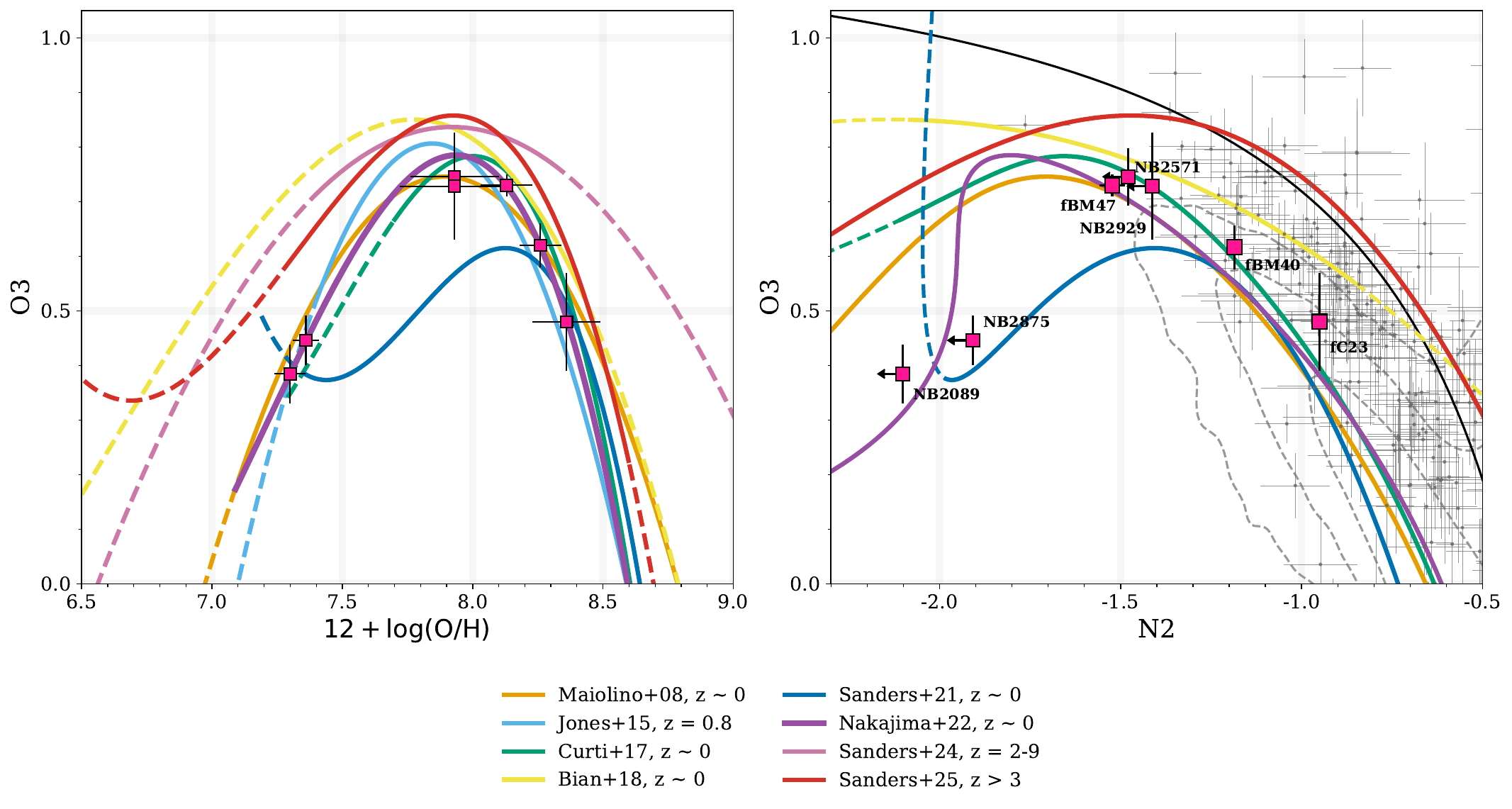}
   \caption{\textit{Left:} O3 vs.\ $12 + \log(\mathrm{O/H})$ diagram with O3-based metallicity calibration curves, as indicated in the legend. While most calibrations agree at intermediate to high metallicities ($12+\log(\mathrm{O/H}) \gtrsim 7.5$), they show substantial variation in O3 at fixed metallicity in the low-metallicity regime. We include the CECILIA Faint points (pink squares) based on their O3 measurements and inferred metallicities from the \cite{Nakajima22} curve. Because the O3–metallicity relation is double-valued, we use N2 (right panel) as a branch discriminator (turnover at $\mathrm{N2} \approx -1.7$). For galaxies with only upper limits in N2 that are consistent with both branches, we take the average of the two inferred metallicities and assign uncertainties that encompass the full possible range. \textit{Right:} N2-BPT diagram \citep{Baldwin1981} showing SDSS $z \sim 0$ galaxies (gray contours), KBSS LBGs \citep{stro2017}, and individual CECILIA Faint galaxies (pink squares; $2\sigma$ limits). The solid black line marks the AGN classification boundary from \citet{Kauffmann2003}. Colored curves show metallicity calibrations for which both O3- and N2-based measurements are available at low and high redshift (same as in the left panel, except \citealt{Jones2015} and \citealt{Sanders2024}). We adopt the \citet{Nakajima22} calibration for this study, which best tracks our data in the N2-BPT space.}
   \label{fig:n2bpt}
\end{figure*}

In the right panel of Figure~\ref{fig:MS}, we show the mass–excitation (MEx) diagram, plotting the log([O\,\textsc{iii}]~$\lambda5007$/H$\beta$) ratio (hereafter O3) versus stellar mass for the same samples as in the left panel. The MEx diagram provides a diagnostic means of distinguishing between star-forming galaxies and AGNs, with the empirical $z \lesssim 1$ demarcation curves from \citet{Juneau2014} shown as dashed purple lines, and their high-redshift ($z\sim2$) counterparts from \citet{Coil2015} indicated by black dashed lines shifted by $\Delta\log M_\ast = 0.75$~dex. All CECILIA Faint galaxies lie below both of these boundaries, consistent with star formation. At $M_\ast \gtrsim 10^{8}\,M_\odot$, the CECILIA Faint galaxies, the \citet{Li2023} dwarfs, and the low-mass tail of the KBSS-LBGs show broadly consistent excitation properties. However, at lower stellar masses, the two faintest CECILIA galaxies show lower O3 ratios than the majority of the \citet{Li2023} sample. This is likely due, at least in part, to the different selection techniques used for the two samples. All the galaxies in the CECILIA Faint sample are selected by their rest-UV colors or Ly$\alpha$ emission. Due to the very deep G235M spectroscopy, all of the galaxies in the KBSS-Ly$\alpha$ sample targeted by  our JWST/NIRSpec observations yielded significant detections of the [O\,\textsc{iii}] and H$\beta$ emission lines, allowing their placement on the MEx diagram. This contrasts with the \citet{Li2023} galaxies, which were identified as  [\ion{O}{3}] emitters in JWST/NRISS slitless spectroscopy without any pre-selection based on their rest-UV properties, with the \citet{Coil2015} curve used to remove AGN contaminants. 
Thus, the different O3 distributions at the lowest masses may reflect differences in selection criteria in addition to potential differences in intrinsic excitation conditions or gas-phase metallicities. We explore the implications of these differences for the MZR in Section~\ref{sec:cosmic_noon}.

\subsection{Gas-phase Metallicity}
\label{sec:gas_phase}

Similar to the analysis of \citet{trem2004} as well as many subsequent studies, our approach relies on strong emission lines, which are commonly observed in star-forming galaxies and serve as useful proxies for gas-phase metallicity. Since direct measurements of the electron temperature (via faint auroral lines such as [O\,\textsc{iii}]~$\lambda4364$) are not possible in many of these galaxies, we rely on strong-line calibrations that relate bright-line ratios to oxygen abundance, either empirically or through photoionization models. 

At $z \sim 2$--3, elevated ionization parameters, high O/Fe ratios, and Fe-poor stellar populations produce harder ionizing spectra and distinct excitation properties compared to local galaxies \citep{stei2016, stro2018, Strom2022, Cullen2019, Cullen2021, Stanton2024}. As a result, local calibrations often fail to reproduce the line ratios of high-redshift systems, particularly at low stellar masses, motivating the use of updated calibrations tailored to early cosmic epochs.

We adopt the O3 ratio as our primary strong-line metallicity diagnostic. This choice is driven by observational constraints: we lack coverage of [O\,\textsc{ii}]~$\lambda3727$ for all sources, preventing the use of the R23 $\equiv$ log$(([\mathrm{OII}]\lambda3727 + [\mathrm{OIII}]\lambda\lambda4960,5007)/\mathrm{H}\beta)$ index, and we measure only upper limits on [N\,\textsc{ii}]~$\lambda6585$ for most targets, making N2-based (\textit{N2} $\equiv$ log$([\mathrm{NII}]\lambda6585/\mathrm{H}\alpha)$) and $\mathrm{O3N2}$-based ($\mathrm{O3N2}$ $\equiv$ O3 $-$ N2) calibrations unreliable. 

\begin{deluxetable*}{lccc}
\tablecaption{Slopes and intercepts for various calibrations}
\tablewidth{\textwidth}
\tablehead{
  \colhead{Calibration} &
  \colhead{$Z_{10}$} &
  \colhead{$\gamma$} &
  \colhead{Calibrated Range (12+log(O/H))}}

\startdata
\cite{Nakajima22}  & $8.58 \pm 0.14$ & $0.48 \pm 0.11$ & 6.9 -- 8.9 \\
\hline
\hline
\cite{Maiolino2008} & $8.47 \pm 0.15$ & $0.46 \pm 0.10$ & 7.1 -- 9.2 \\
\cite{curt2017}     & $8.61 \pm 0.11$ & $0.47 \pm 0.06$ & 7.7 -- 8.8 \\
\cite{Bian2018}     & $9.00 \pm 0.13$ & $0.83 \pm 0.11$ & 7.8 -- 8.4 \\
\cite{Sanders2025}  & $8.34 \pm 0.14$ & $0.44 \pm 0.11$ & 7.3 -- 8.6 \\
\enddata

\tablecomments{Intercepts and slopes for our CECILIA Faint galaxies for different O3-based metallicity calibration curves. Ranges indicate the metallicity validity domains of each calibration.}
\label{tab:slopes_intercepts}
\end{deluxetable*}

A number of O3-based metallicity calibrations have been proposed in the literature, spanning both empirical and theoretical methodologies (Figure~\ref{fig:n2bpt}, left panel). Relations derived by \citet{Jones2015}, \citet{curt2017}, \citet{Sanders2021}, \citet{Sanders2024}, and \cite{Sanders2025} are all based on direct $T_e$ metallicities obtained from stacked or individual galaxy spectra across a range of redshifts ($z \sim 0$--$9$). 
The calibration of \citet{Maiolino2008} combines empirical constraints at low metallicities with photoionization model predictions from \citet{Kewley2002} at higher metallicities, calibrated using SDSS DR4 galaxies. \citet{Bian2018} defines local analogs of $z \sim 2.3$ galaxies based on their location within $\pm0.04$~dex of the high-redshift star-forming sequence on the BPT diagram, specifically requiring that $\mathrm{O3} > \frac{0.67}{N2 - 0.33} + 1.09$, $\mathrm{O3} < \frac{0.67}{N2 - 0.33} + 1.17$, and $\mathrm{N2} < -0.5$ \footnote{The \cite{Bian2018} BPT-based selection aims to ensure that the analogs exhibit ionized ISM conditions similar to those of $z \sim 2.3$ galaxies, with the goal of deriving empirical, $T_e$-based metallicity calibrations applicable to galaxies at $z \sim 2$--5.}. 

Finally, \citet{Nakajima22}, which we select for our analysis as described below, established an empirical relationship between strong-line metallicity diagnostics and direct $T_e$-based metallicities using a sample of local analogs to high-redshift, extremely metal-poor galaxies. These analogs were primarily selected via machine-learning classification from the Subaru/Hyper Suprime-Cam photometric catalog and SDSS, and confirmed spectroscopically via the auroral [O\,\textsc{iii}]~$\lambda4364$ line. The resulting sample spans a wide range of stellar masses ($10^{4}$--$10^{8}\,M_{\odot}$) and star formation rates at $z < 0.05$. To extend the calibration over a broader metallicity range, they supplemented the sample with stacked SDSS galaxy spectra binned by strong-line ratios, excluding AGN using the BPT diagram, resulting in a combined dataset that spans $12 + \log(\mathrm{O}/\mathrm{H}) \approx 6.9$--$8.9$.

The calibrations span an O3 range of approximately $0.3$--$0.8$, with most of them capturing the O3 values of our sources in at least one of their ``branches,'' with the exceptions of \citet{Bian2018}, \citet{Sanders2021}, and \citet{Sanders2024}. While most calibrations largely agree at intermediate to high metallicities ($12+\log(\mathrm{O}/\mathrm{H}) \gtrsim 8.0$), they diverge significantly at lower metallicities, where we expect a substantial fraction of our galaxies to lie. Most relations reproduce the observed turnover in O3 near $12 + \log(\mathrm{O}/\mathrm{H}) \approx 8.0$, with the exception of \citet{Sanders2021}, which peaks at a slightly higher metallicity and exhibits the lowest normalization among the calibrations considered. These differences imply metallicity offsets of up to $\sim0.6$~dex at fixed O3, reflecting systematic uncertainties in the absolute metallicity scale.

We use the published O3- and N2-based metallicity calibrations to define predicted tracks in O3--N2 space as a function of metallicity, and compare these with the positions of our galaxies in the N2-BPT diagram (Figure~\ref{fig:n2bpt}, right panel). Most of the calibrations yield tracks that are overall consistent with our galaxies within the measurement uncertainties. At the lowest $\mathrm{N2}$ values, some calibrations do not reproduce the observed turnover in $\mathrm{O3}$, but our two lowest-$\mathrm{N2}$ sources are upper limits and thus remain consistent with these tracks within errors. The analog-based $T_e$ calibration of \citet{Bian2018} reproduces our galaxies at intermediate $\mathrm{N2}$ but fails to capture the low-$\mathrm{N2}$ turnover \citep[see also][]{Raptis2025}, and therefore we do not adopt it. In contrast, the \citet{Nakajima22} calibration reproduces the locus of our galaxies across the full range of $\mathrm{N2}$ values, including the lowest-$\mathrm{N2}$ sources, and provides the best overall agreement with our data. Given these considerations, we adopt the \citet{Nakajima22} calibration for our metallicity analysis.


Because the O3-based calibration is double-valued, we use the N2 index as a branch discriminator, adopting the turnover point of the \citet{Nakajima22} relation at $\mathrm{N2} \approx -1.7$ as the threshold. For the two sources where N2 is an upper limit and consistent with both branches within uncertainties (NB2571 and NB2929), we report a metallicity range spanning both solutions. For calibration relations whose defined branches do not overlap the observed O3 values within their nominal ranges, we use Monte Carlo sampling of the observed O3 uncertainties to determine whether any realizations intersect the calibration curve, and adopt the corresponding metallicity solutions. This does not arise for the \cite{Nakajima22} calibration. Metallicity uncertainties are estimated using a Monte Carlo approach: for each source we generate $5{,}000$ realizations of the observed O3 ratio, perturbing within the measurement errors, and recompute the metallicity solutions for each draw. The standard deviation of the resulting metallicity distribution provides the $1\sigma$ uncertainty.

The derived oxygen abundances, along with their uncertainties, are listed in Table~\ref{table:properties} and shown in the left panel of Figure~\ref{fig:n2bpt} (pink squares). The measured gas-phase metallicities span the range $12 + \log(\mathrm{O/H}) \approx 7.3$--8.4, with typical uncertainties of $\sim$0.1--0.2~dex per galaxy, based on propagation of line flux errors. These uncertainties do not include the systematic uncertainties associated with the choice of strong-line calibration and thus likely underestimate the true uncertainties in the metallicities.

\begin{figure*}[t]
   \centering
   \includegraphics[width=1\textwidth]{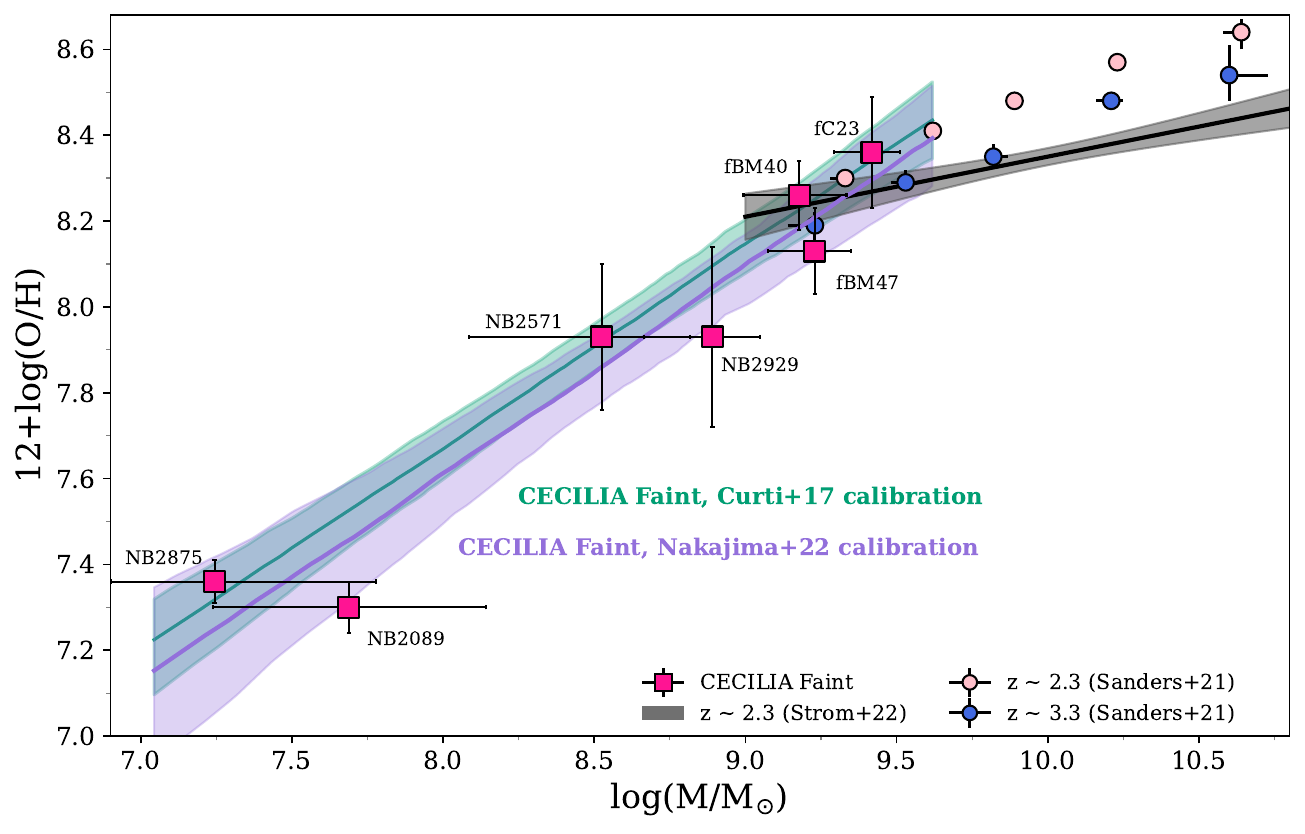}
   \caption{MZR for our CECILIA Faint sample, with gas-phase metallicities $12 + \log(\mathrm{O}/\mathrm{H})$ derived using the O3-based calibration of \citet{Nakajima22}. The solid purple line shows the best-fit linear relation from our MCMC analysis, with the shaded region indicating the $1\sigma$ uncertainty. The slope inferred using the \citet{curt2017} calibration is shown in teal for comparison. Our measurements are compared with stacked spectra at $z \sim 2.3$ and $z \sim 3.3$ from \citet[][pink and blue circles, respectively]{Sanders2021}, and with the KBSS relation determined using photoionization models \citep[][black line and gray shaded region]{Strom2022}. Both MOSDEF and KBSS are consistent with our highest-mass ($M_\star > 10^{9}\,M_\odot$) CECILIA Faint galaxies. The two strong-line calibrations \citep{curt2017, Nakajima22} agree within their $1\sigma$ uncertainties, highlighting that the steeper slope we observe at low masses is not driven by the choice of calibration.}

   \label{fig:mzr}
\end{figure*}

\subsection{Mass-metallicity relation}
\label{mzr}

We find a strong monotonic relationship between stellar mass and oxygen abundance in our sample. The derived oxygen abundances are shown as a function of stellar mass in Figure~\ref{fig:mzr}. We model the MZR using a Bayesian approach implemented with the Markov Chain Monte Carlo (MCMC) sampler \texttt{emcee} \citep{Foreman-Mackey2013} with 50 walkers and 5000 steps. Given the small sample size, we fit a simple linear relation:

\begin{equation}
12 + \log(\mathrm{O}/\mathrm{H}) = Z_{10} + \gamma \cdot \log\left(\frac{M_\star}{10^{10}\,M_\odot}\right),
\end{equation}
where \( Z_{10} \) is the metallicity at \( 10^{10}\,M_\odot \) and \( \gamma \) is the slope.

We incorporated measurement uncertainties on both stellar mass and metallicity, as well as an intrinsic scatter term \(\sigma_{\mathrm{int}}\) to account for additional variance in the metallicity at fixed mass beyond measurement errors. The likelihood function assumes Gaussian uncertainties with total variance \(\sigma_{\mathrm{tot}, i}^2 = \sigma_{\mathrm{meas}, i}^2 + \left(\gamma \times \sigma_{\log M, i}\right)^2 + \sigma_{\mathrm{int}}^2\), where \(\sigma_{\mathrm{meas}, i}\) and \(\sigma_{\log M, i}\) are the metallicity and stellar mass measurement uncertainties for the \(i\)-th galaxy, respectively. For the intrinsic scatter prior, we adopt a Gaussian prior centered on the intrinsic scatter value measured in the KBSS LBG sample, using the same O3-based \citealt{Nakajima22} calibration curve as in our faint sample (see Section \ref{sec:cosmic_noon}), \(\sigma_{\mathrm{int}} = \sqrt{\sigma_{\mathrm{obs}}^2 - \sigma_{\mathrm{meas}}^2} = 0.10 \pm 0.01\) dex. Here, \(\sigma_{\mathrm{obs}}\) is the observed dispersion of metallicity residuals around the error-weighted best-fit MZR in KBSS, and \(\sigma_{\mathrm{meas}}\) quantifies the contribution from measurement errors. Uniform priors were placed on \(Z_{10}\) and \(\gamma\) over physically plausible ranges, specifically \(Z_{10} \in [7.0,\, 9.5]\) and \(\gamma \in [-1.0,\, 1.0]\). The posterior medians and symmetric $1\sigma$ interval define the best-fit parameters and uncertainties, yielding \(Z_{10} = 8.58 \pm 0.14\) and \(\gamma = 0.48 \pm 0.11\), over the stellar mass range \(\log(M_\star/M_\odot) \sim 7.2\)--9.4.

For completeness, we include Table~\ref{tab:slopes_intercepts}, which lists the slopes and intercepts we derive for a variety of O3-based metallicity calibration curves from Figure~\ref{fig:n2bpt}, along with their calibrated metallicity ranges (beyond which extrapolation is required). Each of these MZRs is calculated in the same way as described above, but using a different calibration curve. Among these, we include the calibration of \citet{curt2017} as a reference because it traces the CECILIA Faint galaxies' locations in the N2-BPT space, making it a physically appropriate comparison for the MZR shown in Figure~\ref{fig:mzr}. However, for many of our sources the remaining calibrations require significant extrapolation, which is one of the main reasons we adopt the \citet{Nakajima22} calibration as our baseline throughout this work.

Finally, to constrain the intrinsic scatter, $\sigma_{\mathrm{int}}$, given our limited sample size, we performed a second calculation in which the slope and normalization of the MZR were fixed to the posterior medians from the initial MCMC fit. In contrast to the Gaussian prior used in the MCMC analysis (which incorporates the KBSS measurement as prior information), here we adopt a uniform prior on $\sigma_{\mathrm{int}}$ over the range 0–0.5~dex. This choice ensures that the constraint is driven solely by the CECILIA Faint data, without bias from the more massive KBSS galaxies. The resulting posterior distribution yields a 95\% confidence upper limit of $\sigma_{\mathrm{int}} < 0.22$~dex.

\section{Discussion}
\label{Discussion}

As described in Section~\ref{mzr},our measurements of the low-mass $z\sim2-3$ MZR from the CECILIA Faint sample indicate a  steep slope and a relatively small intrinsic scatter. We also tested for secondary dependencies and find no significant correlation between metallicity and star-formation rate. In the subsequent sections, we compare the low-mass MZR slope at Cosmic Noon to measurements at higher masses and assess its robustness, relate our MZR to other low-mass galaxies at comparable redshifts, and discuss the possible evolution of the MZR with redshift. Given the limitations of our small sample size, we defer a detailed analysis of intrinsic scatter and any dependence on SFR --- including possible connections to the Fundamental Metallicity Relation (FMR; \citealt{Ellison2008, Mannucci2010, Andrews2013}) --- to future work that will incorporate larger statistical samples.

\subsection{Low-mass MZR Slope at Cosmic Noon}
\label{sec:slope_at_cosmic_noon} 
The MZR encodes two key aspects of galaxy evolution: the effectiveness with which galaxies retain the metals they produce and the efficiency with which they convert their gas reservoirs into stars and heavy elements. Competing frameworks emphasize one or the other, but resolving their relative roles requires robust constraints across stellar mass and redshift. Our CECILIA Faint galaxies extend the MZR to $\log M_\star/M_\odot < 7.5$ at $z \sim 2$, providing one of the lowest-mass reference points yet measured at Cosmic Noon.

At $z\sim2$, two major spectroscopic surveys of star-forming galaxies are KBSS \citep{Strom2022} and MOSDEF \citep{Sanders2021}. Both focused on galaxies with $M_\star \gtrsim 10^{9}\,M_\odot$: \cite{Strom2022} derived gas-phase metallicities by fitting rest-optical strong-line ratios with photoionization-model grids to infer $Z_{\rm neb}$ and then converting to $12+\log(\mathrm{O/H})$, while \cite{Sanders2021} obtained metallicities by fitting the observed ratios O32, [O\,III]/H$\beta$, and [Ne\,III]/[O\,II] with the high-redshift-analog calibrations of \citet{Bian2018}. In Figure~\ref{fig:mzr}, we plot the published MZR fit from KBSS (black line with gray shaded region showing the $1\sigma$ uncertainty) and the stacked spectra measurements from MOSDEF at $z \sim 2.3$ and $z \sim 3.3$ (pink and blue circles, respectively).

The three most massive CECILIA Faint galaxies ($M_\star \gtrsim10^{9}\,M_\odot$) lie in agreement with both KBSS and MOSDEF, demonstrating that our adopted calibration is consistent at the high-mass end. However, both KBSS and MOSDEF reported significantly shallower MZR slopes ($\gamma\sim0.14$–0.20), whereas our relation — extending down to $\log(M_\star/M_\odot)\sim7$ — is much steeper, with $\gamma =0.48\pm0.11$, consistent across multiple calibrations (Table~\ref{tab:slopes_intercepts}). 

\begin{figure}[t]
   \centering
   \includegraphics[width=\columnwidth, trim=7 0 0 0, clip]{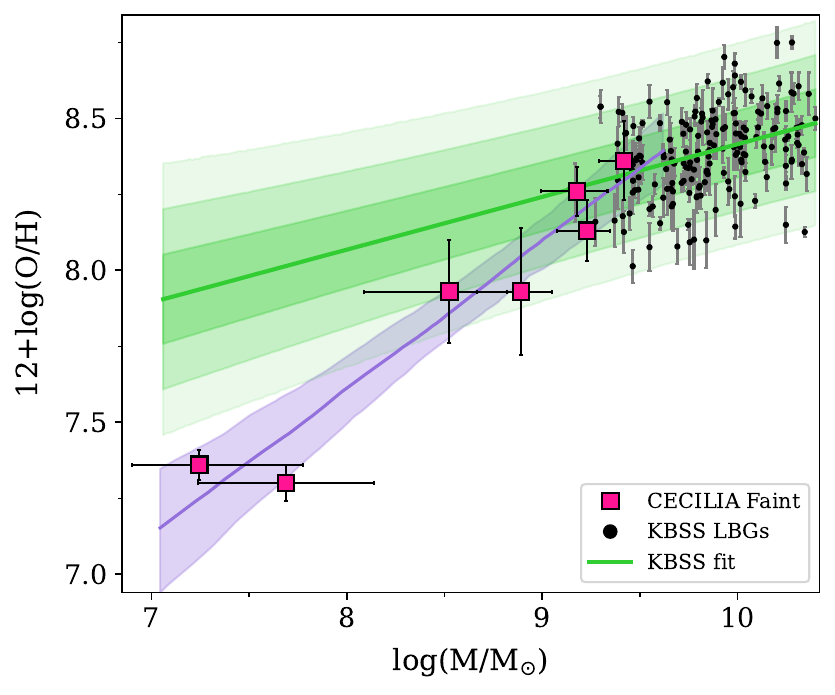}
   \caption{Comparison of the CECILIA Faint galaxy metallicities (pink squares) to the KBSS MZR at $z\sim2$ (black circles), where metallicities were inferred using the \cite{Nakajima22} O3-based calibration for consistency with the CECILIA Faint sample. The solid green line shows the median of 10,000 bootstrap fits to the KBSS sample, and the shaded regions indicate the 1$\sigma$, 2$\sigma$, and 3$\sigma$ prediction regions, from darkest to lightest, derived from the bootstrap ensemble after augmentation with the measured KBSS intrinsic scatter (0.10 dex). The purple line and shaded region show the \citet{Nakajima22} fit to our data and its 1$\sigma$ uncertainty. The two lowest-mass CECILIA Faint galaxies that drive the MZR slope are inconsistent with the KBSS MZR.}
   \label{fig:boot_slope}
\end{figure}

To test whether the difference in slope could arise solely from the adopted abundance calibrations, we re-derived the metallicities of individual KBSS galaxies by applying the same O3-based \citet{Nakajima22} calibration used for our sample to the published O3 measurements \citep{stro2017}; the recalculated KBSS metallicities are presented in Figure~\ref{fig:boot_slope}. We also updated their stellar masses using continuity SFHs from \citet{korh2025}, which are more consistent with the assumptions adopted for the CECILIA Faint stellar mass calculations (Section~\ref{sec:m_sfr}). Using an unweighted fit to these re-derived metallicities, we find an MZR slope of $\gamma = 0.17 \pm 0.03$, measured over the 5–95th percentile in stellar mass (Figure~\ref{fig:boot_slope}). When we instead adopt the constant-SFH stellar masses from \citet{Theios2019} — the same mass scale used by \citet{Strom2022} — while keeping our O3-based metallicities fixed, the slope becomes $\gamma = 0.13 \pm 0.02$. This value is consistent with the $\gamma = 0.14 \pm 0.05$ reported by \citet{Strom2022}, demonstrating that when the same stellar-mass scale is used, our O3-based metallicities reproduce their MZR slope despite the different metallicity estimators. The somewhat steeper slope obtained with the continuity-SFH masses reflects the systematic dependence of the inferred MZR slope on the adopted stellar mass calibration. 

Notably, regardless of the assumptions described above, the KBSS slope remains substantially shallower than that of the CECILIA-Faint MZR. To quantify both the KBSS best-fit MZR and its intrinsic scatter, we drew $N_{\rm boot}=10{,}000$ bootstrap resamples of the KBSS linear fit, evaluated each fit on a common mass grid, and then generated a predictive distribution at each mass by adding MC noise that represents the KBSS intrinsic scatter ($\sigma_{\rm int}=0.10$\,dex): for every bootstrap curve and mass grid point we drew $n_{\rm MC}=200$ independent perturbations from $\mathcal{N}(0,\sigma_{\rm int})$ and added them to the bootstrap-predicted metallicity. From the resulting predictive ensemble we computed the median and the 68\%, 95\%, and 99.7\% bands (Figure~\ref{fig:boot_slope}), and we compared the CECILIA Faint measurements to this distribution. Our lowest-mass galaxies (NB2875 and NB2089) are inconsistent with the extrapolated KBSS relation ($p < 1.3\times10^{-3}$), while the two intermediate-mass systems are marginally consistent with the KBSS distribution, given their relatively large O/H uncertainties. Notably, the faintest galaxies in our sample are Ly$\alpha$-selected, whereas the KBSS and MOSDEF galaxies are continuum-selected. Ly$\alpha$ selection is known to favor lower-metallicity systems at fixed mass \citep[e.g.,][]{Trainor2019}, which could in principle steepen the inferred MZR slope. Considering this effect, we note that our lowest-mass galaxies fall well below the 99.7\% confidence interval of the extrapolated KBSS MZR, inconsistent with the low-metallicity envelope of the KBSS extrapolated slope. This suggests the steepening of the low-mass slope is not solely attributable to selection bias, but at least in part likely an intrinsic property of low-mass galaxies at this epoch.

Given the weak continuum detections of our faintest galaxies, it is also conceivable that the SFR and $A_V$ priors applied in our SED fitting (Section~\ref{sec:m_sfr}) cause us to overestimate the mass of these sources, which could also cause an apparent steepening of the MZR slope. We therefore test the possibility of significantly lower masses by extrapolating the KBSS MZR to the measured metallicities of NB2875 and NB2089. For these galaxies, the KBSS relation would require stellar masses of $\sim 10^{5.5}$ and $\sim 10^{5.3}\,M_\odot$, approximately two orders of magnitude lower than those we infer. Given their H$\alpha$-based star formation rates (Table~\ref{t:galaxies}), such masses would correspond to formation timescales of $\sim 10^{5.6}$ and $\sim 10^{5.1}$ yr — much shorter than the lifetime of massive stars or the timescale for core-collapse supernova feedback. It is therefore extremely unlikely that such rapid mass assembly could have occurred in both of these galaxies, suggesting that the apparent MZR steepness cannot be explained by the method used to estimate the mass of these sources.


To interpret this apparent steepening of the MZR at the low-mass end, it is useful to consider the competing physical pictures of how the slope is set. In the feedback-driven framework, $v_w$ denotes the characteristic velocity of feedback-driven galactic outflows, and the slope is determined by the mass-loading factor ($\eta$) of galactic winds: momentum-driven winds predict $\eta \propto v_w^{-1}$, producing a relatively shallow mass dependence, while energy-driven winds predict $\eta \propto v_w^{-2}$, leading to a steeper slope at the low-mass end. Our steep slope is more consistent with the latter scenario, suggesting that outflows from faint galaxies at $z \sim 2$ are particularly effective at ejecting metals from shallow potential wells. This interpretation aligns with models in which the mass dependence of feedback becomes stronger at earlier epochs \citep[e.g.,][]{Finlator2008, Sanders2023}.

An alternative explanation emphasizes star formation efficiency as the dominant driver of the slope. Low-mass galaxies are generally gas-rich but inefficient at converting gas into stars and metals, and this inefficiency increases toward higher redshift \citep{Hughes2013}. In this view, the steep slope we measure implies that galaxies below $10^{8}\,M_\odot$ at $z \sim 2$ are forming stars with particularly low efficiency, delaying their chemical enrichment relative to more massive systems. We note that this low efficiency may itself be a consequence of strong stellar feedback, which can expel gas and suppress star formation in low-mass galaxies. Our results therefore do not uniquely favor one mechanism over the other, but they highlight the importance of considering both feedback and efficiency as regulators of the MZR.

We caution that slope measurements are sensitive to systematic uncertainties in metallicity calibrations, with differences of $\Delta \gamma \sim 0.4$ possible depending on the diagnostic used (Table \ref{tab:slopes_intercepts}), although we also emphasize that the alternative calibrations yield even steeper slopes than \cite{Nakajima22}. Nevertheless, our comparison with the KBSS sample and the robustness tests with our faintest galaxies indicate that the steepening reflects that low-mass galaxies have stronger mass-dependent feedback and/or less efficient star formation at $z \sim 2$.

\subsection{Comparison with Other Low-Mass Cosmic Noon Samples and Metallicity Diagnostics}
\label{sec:cosmic_noon} 
We next compare our results with other low-mass samples at similar redshift. Because O3 is sensitive not only to metallicity but also to ionization conditions \citep{stro2017, stro2018}, it is essential to compare our results with metallicities inferred at similar redshift using alternative methods. In this section, we therefore consider results based on other strong-line indicators as well as direct $T_e$ measurements.

\begin{figure}[t]
   \centering
   \includegraphics[width=\columnwidth, trim=7 0 0 0, clip]{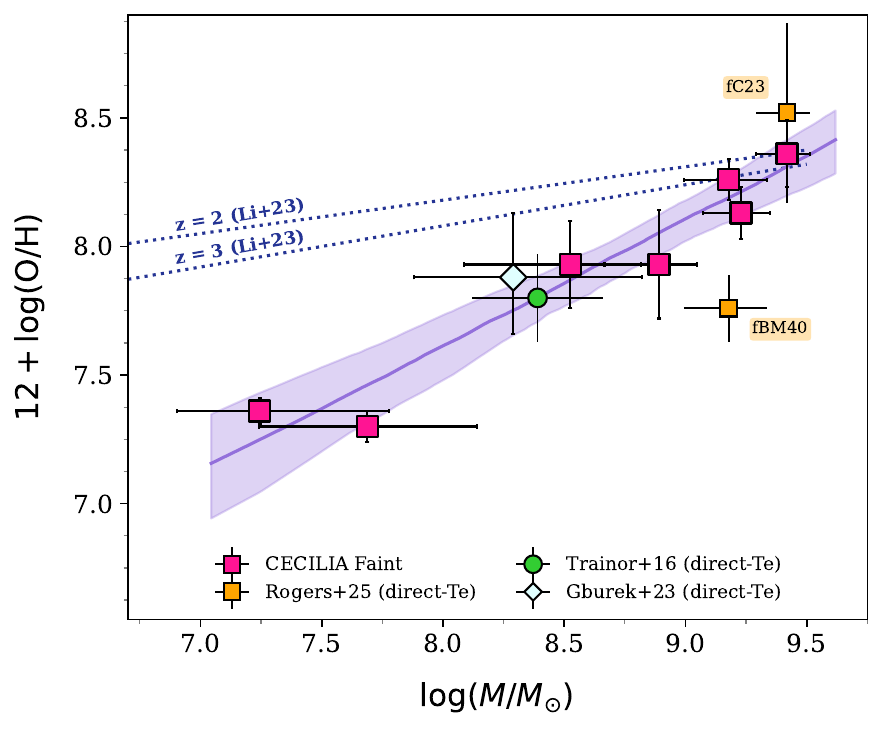}
   \caption{A comparison of various low-mass MZR measurements at Cosmic Noon. The MZR for our CECILIA Faint sample (pink squares) is shown with the best-fit relation and scatter (purple line and shaded region). We include other direct-\(T_e\) metallicity samples: two CECILIA Faint sources from \citet[][orange squares]{Rogers2025}, a stack of 52 LAEs at $z \sim 2.56$ from \citet[][green circle]{Trainor2016}, and a stack of gravitationally lensed dwarfs at $z \sim 2.3$ from \citet[][aqua diamond]{Gburek2023}, which appear generally consistent with the CECILIA Faint sample.
   We also show the MZR fits at $z \sim 2$ and $z \sim 3$ from \citet[][blue dotted lines]{Li2023}. As discussed in the text, we expect that significant differences in the selection of galaxies between our sample and the \citet{Li2023} sample drive the differences in our inferred MZRs.}
   \label{fig:comparison_to_li23}
\end{figure}

\begin{figure*}[t]
   \centering
   \includegraphics[width=1\textwidth]{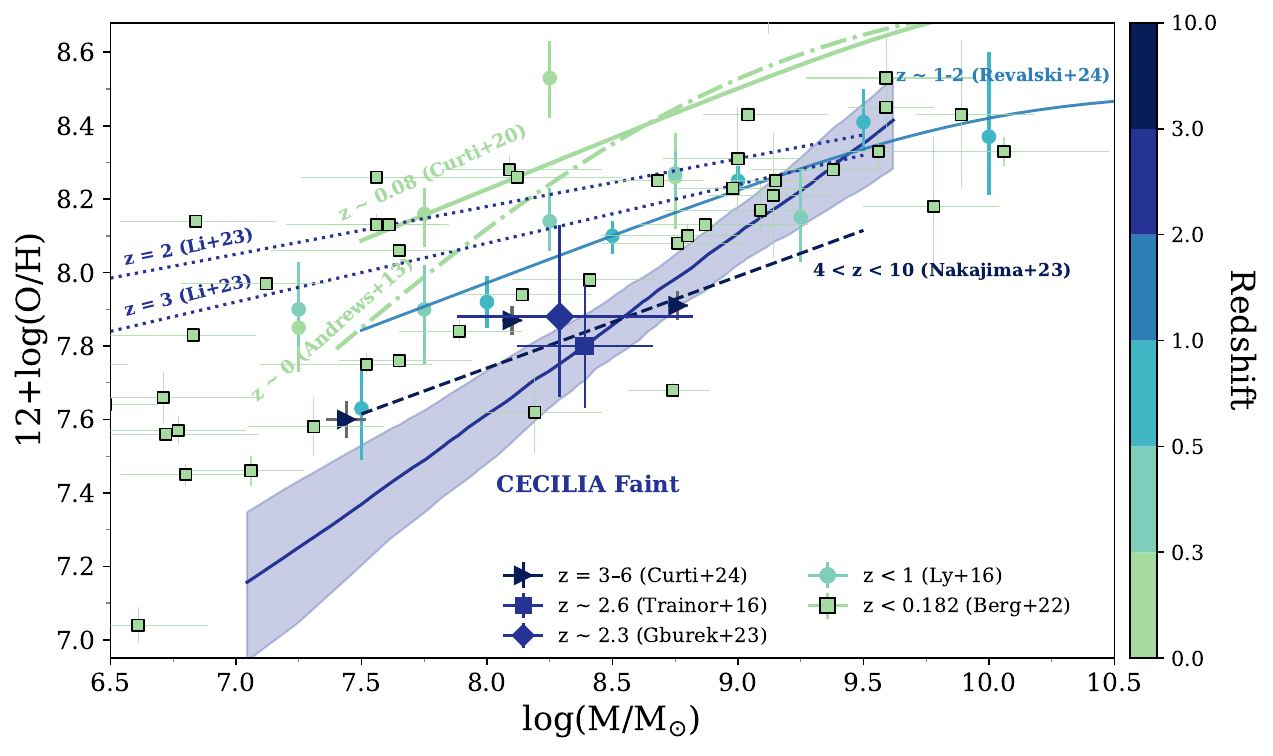}
   \caption{Redshift evolution of the low-mass MZR. At low redshift, we show the SDSS-based relations of \citet[][$z \sim 0$, dash–dotted line]{Andrews2013} and \citet[][$z \sim 0.08$, solid line]{Curti2020}, together with CLASSY galaxies ($z < 0.18$; squares). Individual galaxies from the MACT survey \citep[][circles]{Ly2016} trace the decline to $z < 1$, while the relation from \citet[][solid line]{Revalski2024} extends this trend at $z \sim 1$–2. At redshifts comparable to our sample, we include the stack at $z \sim 2.6$ from \citet[][square]{Trainor2016}, the lensed dwarf stack of \citet[][diamond]{Gburek2023}, and the MZR fits at $z \sim 2$ and $z \sim 3$ from \citet[][dotted lines]{Li2023}. At higher redshift ($z > 3$), we show stacks from \citet[][triangles]{Curti2024} and the CEERS/NIRSpec fit of \citet[][dashed line]{naka2023}. Colors indicate the average redshift of each sample, with green corresponding to the lowest redshifts and dark blue to the highest. Overall, metallicity at fixed stellar mass decreases with redshift, reflecting less mature chemical enrichment, while the MZR slope shows little evolution. Past Cosmic Noon, there is no apparent evolution in the normalization of the MZR.}

   \label{fig:mzr_evol}
\end{figure*}

Figures~\ref{fig:comparison_to_li23} and~\ref{fig:mzr_evol} together place our CECILIA Faint galaxies in the broader context of low-mass MZR measurements. Here we focus on the $z \sim 2$ comparisons, while a detailed discussion of the redshift evolution across $0 < z < 10$ is presented in Section~\ref{sec:redshift_evolution}.

In Figure~\ref{fig:comparison_to_li23}, we compare our O3–based metallicities with recent direct–$T_e$ measurements of low-mass galaxies at $z\sim2-3$, including values from \citet{Rogers2025} for two of the faint LBGs in our CECILIA Faint sample (orange squares).
For fC23, the direct metallicity agrees within $1\sigma$ of our O3 estimate, whereas for fBM40 the direct value is substantially lower.
We also include the \citet{Trainor2016} stack of 52 faint LAEs from the KBSS–Ly$\alpha$ survey at $z \approx 2.56$ (green circle), for which an average metallicity was derived using the direct–$T_e$ method and an average stellar mass computed in the same way as for our sample (Section~\ref{sec:m_sfr}). Overall, the \citet{Trainor2016} stack is in excellent agreement with the CECILIA faint measurements.

Another complementary low-mass benchmark at $z \approx 2.3$ is provided in Figure~\ref{fig:comparison_to_li23} by \citet{Gburek2023}, who stack 16 gravitationally lensed dwarfs (median $\log M_\star/M_\odot \approx 8.3$) and obtain a direct $T_e$ metallicity of $12+\log({\rm O/H}) \approx 7.9$ from [O\,III]~$\lambda4364$. This value agrees well within the uncertainties with the MZR fit to our CECILIA Faint sample. We also recalculate their metallicity using the \citet{Nakajima22} calibration. Because the available N2 measurement is a $2\sigma$ upper limit (N2 $<-1.3$), we adopt the average metallicity of the two branches, with uncertainties spanning the full allowed range, and find $12+\log({\rm O/H}) \approx 7.9 \pm 0.1$, consistent with the direct determination. These comparisons provide a direct test of our calibration against $T_e$–based measurements and highlight both their overall consistency and the limited samples of $T_e$ measurements at low masses.

Finally, we consider the relation from \citet{Li2023}, who investigated the low-mass ($\log M_\star/M_\odot \sim 6.5$–9.5) MZR with JWST/NIRISS slitless spectroscopy. As described in Section~\ref{sec:m_sfr}, \citet{Li2023} identified lensed dwarf galaxies on the basis of their [\ion{O}{3}] emission and derived metallicities using the \citet{Bian2018} strong-line calibrations applied to O32\footnote{O32 $\equiv \log$([\ion{O}{3}]$\lambda\lambda4960,5007$/[\ion{O}{2}]$\lambda\lambda$3727,3729)} (and O3 when [O\,II] was unavailable). As shown in Figure~\ref{fig:comparison_to_li23}, their reported metallicities lie significantly above both the CECILIA Faint systems and the $T_e$ samples at $M_*\lesssim10^9\,\textrm{M}_\odot$, and their derived low-mass MZR slopes at both $z\sim2$ ($\gamma = 0.13$) and $z\sim3$ ($\gamma =0.16$) are substantially shallower than our relation.

As we do not have [O\,II] detections for most CECILIA Faint galaxies, we cannot directly compare to the \citet{Li2023} results using a matched O32 indicator. 
Similarly, \citet{Li2023} do not have N2 coverage and so cannot distinguish between the high- and low-metallicity branches of the \citet{Nakajima22} calibration, as we have done in this work. To assess the potential impact of the different indicator choices, we estimated the metallicities of the \citet{Li2023} individual galaxies under multiple assumptions of the low or high-metallicity \citet{Nakajima22} O3 branches. If their galaxies with $\log(M/M_*)\lesssim 8.5$ tend to lie on the low-O/H O3 branch and the higher-mass galaxies tend to lie on the high-O/H branch---as is the case for our CECILIA Faint sample---then the MZR slope inferred from the \citet{Li2023} sample using the \citet{Nakajima22} O3 calibration would be steeper than the slope they infer using the \citet{Bian2018} O32 calibration. Regardless of the choice of branch, however, we consistently find that the slopes inferred from the \citet{Li2023} sample are shallower than the relation we find for the CECILIA Faint sample, indicating that differences in calibration alone cannot account for the distinct shapes of the two MZRs.

Instead, it is likely that differences in the selection between the \citet{Li2023} sample and our CECILIA Faint galaxies significantly contribute to the slope difference. Our lowest-mass galaxies are Ly$\alpha$-selected; as we discuss in Section \ref{sec:slope_at_cosmic_noon}, this may bias our sample against higher metallicity galaxies at low masses and thus steepen the MZR slope we measure. Conversely, the \citet{Li2023} galaxy selection using [O \textsc{iii}] emission may be biased against the lowest-metallicity galaxies, whose lower oxygen abundances result in lower [O \textsc{iii}]~$\lambda5007$ line fluxes relative to the Balmer emission lines. Depending on the galaxies' SFRs and lensing magnifications, many low-metallicity systems could lie below the [O \textsc{iii}] flux limit of their survey.  Furthermore, note that due to their strong [O \textsc{iii}]~$\lambda5007$ emission, some fraction of the \citet{Li2023} sources may retain weak AGN contamination that is not fully removed by the \citet{Coil2015} MEx diagnostic (Section~\ref{sec:m_sfr}). Indeed, 60\% of the \citet{Li2023} galaxies have O3 ratios exceeding the peak value of the Nakajima relation, and many of their O3 ratios exceed the maximum values for all O3-based calibrations considered in this work, including \citet{Bian2018}.



These uncertainties underscore the need for larger and deeper spectroscopic samples with full wavelength coverage, multiple independent diagnostics, and a range of selection criteria to robustly establish the low-mass end of the MZR at $z\sim2$--3.

\subsection{Redshift Evolution of the low-mass MZR}
\label{sec:redshift_evolution}

In this section, we examine the redshift evolution of metallicity among low-mass galaxies at fixed stellar mass. This redshift evolution generally reflects lower degrees of chemical enrichment at earlier times. In Figure~\ref{fig:mzr_evol}, we compare our CECILIA Faint MZR to measurements at both lower and higher redshift, which are introduced below if not already discussed in Section~\ref{sec:cosmic_noon}.

Starting from the nearby universe, SDSS-based studies \citep{Andrews2013, Curti2020} report the highest metallicities at fixed stellar mass, consistent with chemically mature systems after billions of years of enrichment. The CLASSY survey \citep{Berg2022} complements these results with 45 star-forming galaxies at $0.002 < z < 0.182$ spanning $6.2 < \log(M_\star/M_\odot) < 10.1$. Their direct-method metallicities ($7.0 < 12+\log(\mathrm{O/H}) < 8.8$), derived from ionic abundances and $T_e$-based diagnostics, show large scatter ($\sigma_{\rm int} = 0.42 \pm 0.07$), but many are consistent with our relation, suggesting that some may serve as useful low-$z$ analogs to faint high-redshift systems. Below $z \sim 1$, the sample of \citet{Ly2016} shows a systematic decline in metallicity with increasing redshift, as traced by their three redshift bins ($z<0.3$, $0.3<z<0.5$, and $0.5<z<1$). At intermediate redshifts ($1 \lesssim z \lesssim 2$), studies such as \citet{Revalski2024} probe down to $\log(M_\star/M_\odot) \approx 7.5$, finding metallicities of $12 + \log(\mathrm{O/H}) \approx 7.8$.

By $z \sim 2$, our CECILIA Faint galaxies provide a new anchor at the low-mass end, extending down to $\log(M_\star/M_\odot) \sim 7$. At $\log(M_\star/M_\odot) \sim 8.2$, the stack of lensed dwarf galaxies from \citet{Gburek2023} is consistent with our MZR as explained in Section \ref{sec:cosmic_noon}. At fixed stellar mass, our metallicities lie $\sim0.5$ dex below the relation reported by \citet{Li2023}; this discrepancy has already been discussed in Section~\ref{sec:cosmic_noon}.

At $z > 3$, the NIRSpec/CEERS analysis of \citet{naka2023} derived an MZR over $z \sim 4$–10 using the O32 index (and O3 when [O\,II] was unavailable), applying the EW(H$\beta$)-dependent prescriptions of \citet{Nakajima22} to account for ionization state and break the R23 degeneracy. \citet{Curti2024} analyzed $z \sim 3$–10 galaxies from JADES, deriving metallicities by combining multiple strong-line diagnostics (primarily ratios such as O3, O32, and $\hat{R}$) against revised empirical calibrations tied to the direct-$T_e$ scale. Both CEERS \citep{naka2023} and JADES \citep{Curti2024} stacks lie well below the relations measured for low-$z$ systems, but they are consistent with the MZR measured from the CECILIA Faint galaxies. The overall trend indicates minimal evolution between the EoR and Cosmic Noon, implying that Cosmic Noon galaxies may serve as useful analogs for EoR systems. 

This result agrees with \citet{Marszewski2025}, who used FIRE-2 simulations which showed that the high-redshift MZR does not evolve over $5 \lesssim z \lesssim 12$. While the \cite{Marszewski2025} simulations do not extend down to the redshift range of our sample, the physical mechanism they identify for the lack of evolution at $z \gtrsim 5$ may similarly help explain the consistency we observe between the MZR we measure at $z \sim 2$--3 and the higher-redshift samples of \cite{naka2023} and \cite{Curti2024}. In their simulations, decreasing redshift leads to more enriched inflows at fixed stellar mass due to increased wind recycling, while the metal mass produced per unit inflowing gas decreases because of reduced star formation efficiency within burst cycles. These two events largely cancel, yielding a nearly invariant MZR across these redshifts.

Taken together, these comparisons show a clear progression: galaxies at fixed mass are most enriched at $z \sim 0$, exhibit strong evolution back to $z \sim 2$, and then remain broadly consistent out to higher redshifts. Our findings at the low-mass end support this overall trend, with comparatively little metallicity evolution beyond $z \sim 2$.

\section{Conclusions}
\label{Conclusions}

We have analyzed ultra-deep JWST/NIRSpec spectroscopy of seven continuum-faint star-forming galaxies at $z\sim2$–3 from the CECILIA survey to investigate the low-mass end of the mass–metallicity relation during Cosmic Noon. Our main conclusions are as follows:

\begin{enumerate}
    
    \item We adopt the O3-based strong-line metallicity calibration of \cite{Nakajima22}, which best traces the full CECILIA Faint sample in N2-BPT space, finding $12 + \log(\mathrm{O/H}) \approx 7.3$--8.4 over $\log(M_\star/M_\odot) \sim 7.3$--9.7. A linear fit to the MZR yields 
    $12 + \log(\mathrm{O/H}) = (8.58 \pm 0.14) + (0.48 \pm 0.11) \cdot \log(M_\star / 10^{10}\,M_\odot)$. Other tested calibrations are generally consistent within $1\sigma$. \textbf{Tables~\ref{table:properties}--\ref{tab:slopes_intercepts}}; \textbf{Figure~\ref{fig:n2bpt}}; \textbf{Sections~\ref{sec:gas_phase}--\ref{mzr}}.

    \item Compared to the higher-mass KBSS relation at $z\sim2$ (computed with the same O3-based calibration) and extrapolated to our lowest-mass regime, our faint galaxies show a significantly steeper trend that cannot be explained by stellar mass uncertainties. This departure reflects a significant steepening of the MZR at the faint end, consistent with either stronger mass-dependent feedback or inefficient star formation. \textbf{Figures~\ref{fig:mzr}--\ref{fig:boot_slope}}; \textbf{Section~\ref{sec:slope_at_cosmic_noon}}.

    \item Our O3-based metallicities generally agree with other low-mass galaxies at $z\sim2$, including both strong-line and direct-$T_e$ measurements, though some discrepancies exist due to calibration choice, branch assignment, or selection effects. The availability of N2 measurements for our CECILIA Faint galaxies allows us to break branch degeneracies, providing a more reliable placement in the low-mass MZR. These comparisons highlight the need for additional low-mass galaxies with multiple independent metallicity diagnostics to robustly establish the MZR at Cosmic Noon. \textbf{Figures~\ref{fig:mzr}, \ref{fig:comparison_to_li23}--\ref{fig:mzr_evol}}; \textbf{Section~\ref{sec:cosmic_noon}}.

    \item We find clear evolution of metallicity at fixed stellar mass: galaxies in the local universe show the highest enrichment, intermediate-$z$ systems have lower metallicities, and our faint $z \sim 2$ galaxies have chemistries comparable to those measured at the epoch of reionization, up to $z \sim 10$. This result suggests that faint systems at $z \sim 2$ may serve as useful analogs of the EoR. \textbf{Figure~\ref{fig:mzr_evol}}; \textbf{Section~\ref{sec:redshift_evolution}}.


\end{enumerate}

Robustly constraining the mass–metallicity relation at Cosmic Noon, especially at low masses, requires larger spectroscopic samples with direct $T_e$–based abundances.
As our comparison with higher-$z$ galaxies has shown, our CECILIA Faint galaxies are likely close analogs of early reionization-era systems, making it imperative to invest JWST time in deep rest-optical spectroscopy to detect the weak auroral lines needed to calibrate strong-line diagnostics.
Such observations will enable consistent cross-calibration of metallicity diagnostics and clarify how feedback and star-formation efficiency shape the MZR, firmly linking faint Cosmic-Noon galaxies to their high-redshift counterparts.

\begin{acknowledgments}

M. R. acknowledges partial support from the Carnegie Institution for Science and the Carnegie Astrophysics Summer Student Internship Program. R.F.T, A.L.S., and G.C.R. acknowledge partial support from the JWST-GO-02593.008-A, JWST-GO-02593.004-A, and JWST-GO-02593.006-A grants, respectively.
N.S.J.R. was also supported by JWST-GO-02593.004-A. These funds were provided by NASA through a grant from the Space Telescope Science Institute, which is operated by the Association of Universities for Research in Astronomy, Inc., under NASA contract NAS 5-03127.
R.F.T. also acknowledges support from the Pittsburgh Foundation (grant ID UN2021-121482) and the Research Corporation for Scientific Advancement (Cottrell Scholar Award, grant ID 28289). ALS acknowledges support from the David and Lucile Packard Foundation. Effort by ALS was completed in part at the Aspen Center for Physics, which is supported by National Science Foundation grant PHY-2210452.
M.V.M. is supported by the National Science Foundation via AAG grant 2205519.

This work is primarily based on observations made with NASA/ESA/CSA JWST, associated with PID 2593, which can be accessed via doi:\dataset[10.17909/x66z-p144]{https://doi.org/10.17909/x66z-p144}. The data were obtained from the Mikulski Archive for Space Telescopes (MAST) at the Space Telescope Science Institute, which is operated by the Association of Universities for Research in Astronomy, Inc., under NASA contract NAS 5-03127 for JWST.

This work is based in part on observations made with the Spitzer Space Telescope, which was operated by the Jet Propulsion Laboratory, California Institute of Technology, under a contract with NASA.

\end{acknowledgments}

\appendix

\section{Dynamical Mass Estimates}
\label{appendix:dynamical_masses}

\begin{figure}[h]
   \centering
   \includegraphics[width=0.5\columnwidth, trim=7 0 0 0, clip]{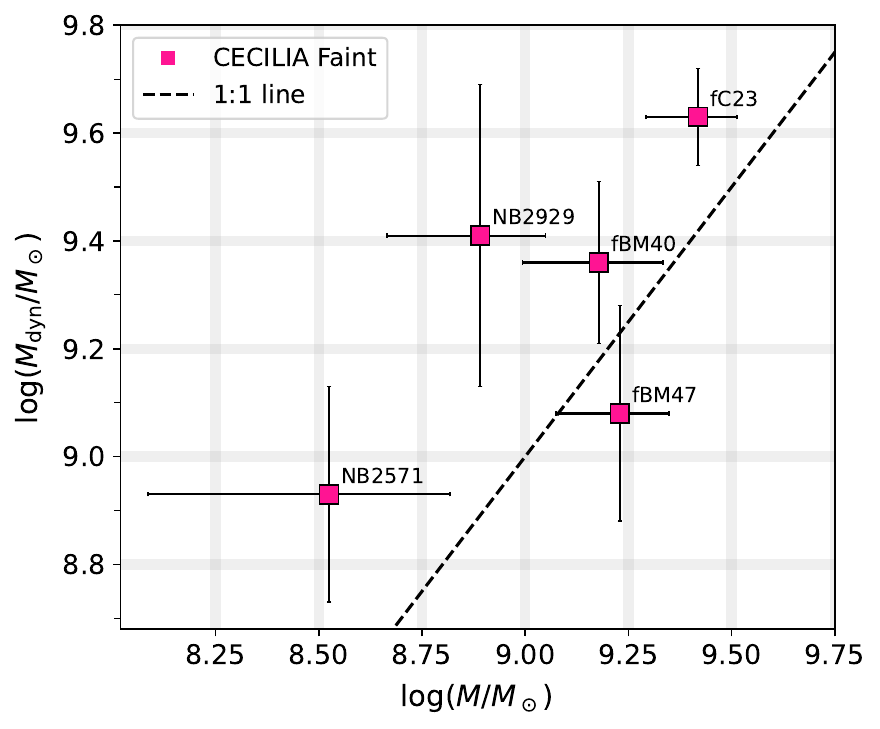}
    \caption{Dynamical vs. stellar masses for the CECILIA Faint sample (pink squares) with the 1:1 line (black dashed). Dynamical masses are computed using the geometric mean of the spatial-profile–based and spectral-resolution–based size estimates to determine the effective radius of each object, as described in the text. Overall, the galaxies show good agreement between their dynamical and stellar masses.}

   \label{fig:stellar_dyn}
\end{figure}

The dynamical masses discussed in Section~\ref{sec:m_sfr} are estimated using the dispersion-based virial estimator of \citet{Erb2006}, $M_{\rm dyn} = C\,\sigma^{2} R / G$, where $\sigma$ is the H$\alpha$ velocity dispersion obtained from MOSFIRE after subtracting the instrumental resolution of $\sim35~\mathrm{km\,s^{-1}}$ \citep{mclean2012} in quadrature to estimate the intrinsic width. The constant $C$ is adopted as in \citet{Erb2006} ($C \approx 3.42$), and $R$ is the galaxy radius, determined through multiple complementary methods as follows.

We begin with the JWST/NIRSpec 2D spectrogram of each source. To measure the spatial extent of the H$\alpha$ emission, we collapse the spectrum over a narrow wavelength window centered on H$\alpha$ (i.e., along the dispersion axis) to obtain a 1D spatial profile. This profile is fit with a Gaussian, and the FWHM is measured. The intrinsic source size is recovered by deconvolving the instrumental PSF, assuming a PSF FWHM of $\sim0.08\arcsec$ for JWST/NIRSpec at 2.5~$\mu$m \citep{Jakobsen2022}. The deconvolved FWHM is converted to a physical scale using the angular scale from our adopted cosmology, and the H$\alpha$-based radius is defined as half of this physical size. Uncertainties are estimated via Monte Carlo simulations, adding Gaussian noise to the profile (with standard deviation based on the pipeline-produced flux uncertainties) and remeasuring the FWHM multiple times to construct a distribution.

In parallel, we also infer galaxy sizes from their spectral resolution.  We begin with the spectral resolution as a function of wavelength for the G235M grating from the NIRSPEC JDox. This resolution is defined for a uniformly illuminated, slit-filling source with width $0.2\arcsec$ (the physical width of the MSA shutters), so unresolved or partially resolved sources with intrinsically-narrow emission lines will have higher effective resolution than this reference case. For the LAEs in our sample, we measure a typical spectral resolution $\sim40\%$ higher than the JDox prediction, which falls between the uniform-slit-filling assumption of JDox and the point-source resolution estimates of \citet{deGraaff2024}. 
For sources smaller than the slit width, the physical size can therefore be inferred by comparing the observed spectral resolution to this reference. We estimate the instrumental resolution using the emission lines H$\alpha$, H$\beta$, [O~III]~$\lambda 4960$, [O~III]~$\lambda 5007$, [N~II]~$\lambda 6585$, and [S~II]~$\lambda\lambda 6718, 6733$, when available.  
For each line, we calculate the observed FWHM and plot it as a function of wavelength, which we then fit with a linear relation.  
A multiplicative factor, $f$, is defined as the ratio required to match our best-fit line to the  FWHM($\lambda$) relation predicted by the JWST/NIRSpec resolution curve \citep{JDox}.  
We then scale the inferred source size as $0.2\arcsec / f$, where $0.2\arcsec$ represents the shutter width and dividing by $f$ corrects the measured resolution to the effective angular size of the source relative to this limit.  
To propagate uncertainties, we recompute the best-fit $f$ using the ratio between the observed and predicted spectral FWHM for each emission line of a given sources and adopt the standard deviation as the uncertainty on the source size. Again, the corresponding radius is taken as half the inferred FWHM.

The spatial- and spectral-resolution-based radii are oriented approximately $90^\circ$ apart, and we therefore adopt their geometric mean as the radius of each galaxy. We further compare these radii to those obtained by fitting the HST/F140W images of our targets with two-dimensional PSF-convolved Gaussian profiles, from which we again compute the geometric mean radius. For the CECILIA Faint galaxies that are well detected in the HST images, the two measurements agree and scatter around the 1:1 relation with an rms scatter of 0.03\arcsec, supporting the use of the averaged radius from the spatial-profile and spectral-resolution methods in the virial equation 
to compute the dynamical mass. The comparison in Figure~\ref{fig:stellar_dyn} shows that the stellar and dynamical mass estimates are in reasonable agreement given the uncertainties.

\bibliographystyle{aasjournal}
\bibliography{aggregate_refs}

@ARTICLE{aspl2021,
       author = {{Asplund}, M. and {Amarsi}, A.~M. and {Grevesse}, N.},
        title = "{The chemical make-up of the Sun: A 2020 vision}",
      journal = {\aap},
         year = 2021,
        month = sep,
       volume = {653},
          eid = {A141},
        pages = {A141},
          doi = {10.1051/0004-6361/202140445},
archivePrefix = {arXiv},
       eprint = {2105.01661},
 primaryClass = {astro-ph.SR},
       adsurl = {https://ui.adsabs.harvard.edu/abs/2021A&A...653A.141A}
}

@ARTICLE{berg2022,
       author = {{Berg}, Danielle A. and {James}, Bethan L. and {King}, Teagan and {McDonald}, Meaghan and {Chen}, Zuyi and {Chisholm}, John and {Heckman}, Timothy and {Martin}, Crystal L. and {Stark}, Dan P. and {Aloisi}, Alessandra and {Amor{\'\i}n}, Ricardo O. and {Arellano-C{\'o}rdova}, Karla Z. and {Bayliss}, Matthew and {Bordoloi}, Rongmon and {Brinchmann}, Jarle and {Charlot}, St{\'e}phane and {Chevallard}, Jacopo and {Clark}, Ilyse and {Erb}, Dawn K. and {Feltre}, Anna and {Gronke}, Max and {Hayes}, Matthew and {Henry}, Alaina and {Hernandez}, Svea and {Jaskot}, Anne and {Jones}, Tucker and {Kewley}, Lisa J. and {Kumari}, Nimisha and {Leitherer}, Claus and {Llerena}, Mario and {Maseda}, Michael and {Mingozzi}, Matilde and {Nanayakkara}, Themiya and {Ouchi}, Masami and {Plat}, Adele and {Pogge}, Richard W. and {Ravindranath}, Swara and {Rigby}, Jane R. and {Sanders}, Ryan and {Scarlata}, Claudia and {Senchyna}, Peter and {Skillman}, Evan D. and {Steidel}, Charles C. and {Strom}, Allison L. and {Sugahara}, Yuma and {Wilkins}, Stephen M. and {Wofford}, Aida and {Xu}, Xinfeng and {Classy Team}},
        title = "{The COS Legacy Archive Spectroscopy Survey (CLASSY) Treasury Atlas}",
      journal = {\apjs},
         year = 2022,
        month = aug,
       volume = {261},
       number = {2},
          eid = {31},
        pages = {31},
          doi = {10.3847/1538-4365/ac6c03},
archivePrefix = {arXiv},
       eprint = {2203.07357},
 primaryClass = {astro-ph.GA},
       adsurl = {https://ui.adsabs.harvard.edu/abs/2022ApJS..261...31B}
}

@ARTICLE{curt2017,
       author = {{Curti}, M. and {Cresci}, G. and {Mannucci}, F. and {Marconi}, A. and {Maiolino}, R. and {Esposito}, S.},
        title = "{New fully empirical calibrations of strong-line metallicity indicators in star-forming galaxies}",
      journal = {\mnras},
         year = 2017,
        month = feb,
       volume = {465},
       number = {2},
        pages = {1384-1400},
          doi = {10.1093/mnras/stw2766},
archivePrefix = {arXiv},
       eprint = {1610.06939},
 primaryClass = {astro-ph.GA},
       adsurl = {https://ui.adsabs.harvard.edu/abs/2017MNRAS.465.1384C}
}

@ARTICLE{erb2006,
       author = {{Erb}, Dawn K. and {Shapley}, Alice E. and {Pettini}, Max and {Steidel}, Charles C. and {Reddy}, Naveen A. and {Adelberger}, Kurt L.},
        title = "{The Mass-Metallicity Relation at z>\raisebox{-0.5ex}\textasciitilde2}",
      journal = {\apj},
         year = 2006,
        month = jun,
       volume = {644},
       number = {2},
        pages = {813-828},
          doi = {10.1086/503623},
archivePrefix = {arXiv},
       eprint = {astro-ph/0602473},
 primaryClass = {astro-ph},
       adsurl = {https://ui.adsabs.harvard.edu/abs/2006ApJ...644..813E}
}

@ARTICLE{ferl2013,
       author = {{Ferland}, G.~J. and {Porter}, R.~L. and {van Hoof}, P.~A.~M. and {Williams}, R.~J.~R. and {Abel}, N.~P. and {Lykins}, M.~L. and {Shaw}, G. and {Henney}, W.~J. and {Stancil}, P.~C.},
        title = "{The 2013 Release of Cloudy}",
      journal = {\rmxaa},
         year = 2013,
        month = apr,
       volume = {49},
        pages = {137-163},
archivePrefix = {arXiv},
       eprint = {1302.4485},
 primaryClass = {astro-ph.GA},
       adsurl = {https://ui.adsabs.harvard.edu/abs/2013RMxAA..49..137F}
}

@ARTICLE{mous2010,
       author = {{Moustakas}, John and {Kennicutt}, Robert C., Jr. and {Tremonti}, Christy A. and {Dale}, Daniel A. and {Smith}, John-David T. and {Calzetti}, Daniela},
        title = "{Optical Spectroscopy and Nebular Oxygen Abundances of the Spitzer/SINGS Galaxies}",
      journal = {\apjs},
         year = 2010,
        month = oct,
       volume = {190},
       number = {2},
        pages = {233-266},
          doi = {10.1088/0067-0049/190/2/233},
archivePrefix = {arXiv},
       eprint = {1007.4547},
 primaryClass = {astro-ph.CO},
       adsurl = {https://ui.adsabs.harvard.edu/abs/2010ApJS..190..233M}
}

@ARTICLE{naka2023,
       author = {{Nakajima}, Kimihiko and {Ouchi}, Masami and {Isobe}, Yuki and {Harikane}, Yuichi and {Zhang}, Yechi and {Ono}, Yoshiaki and {Umeda}, Hiroya and {Oguri}, Masamune},
        title = "{JWST Census for the Mass-Metallicity Star-Formation Relations at z=4-10 with the Self-Consistent Flux Calibration and the Proper Metallicity Calibrators}",
      journal = {arXiv e-prints},
         year = 2023,
        month = jan,
          eid = {arXiv:2301.12825},
        pages = {arXiv:2301.12825},
          doi = {10.48550/arXiv.2301.12825},
archivePrefix = {arXiv},
       eprint = {2301.12825},
 primaryClass = {astro-ph.GA},
       adsurl = {https://ui.adsabs.harvard.edu/abs/2023arXiv230112825N}
}

@ARTICLE{rudi2012,
       author = {{Rudie}, Gwen C. and {Steidel}, Charles C. and {Trainor}, Ryan F. and {Rakic}, Olivera and {Bogosavljevi{\'c}}, Milan and {Pettini}, Max and {Reddy}, Naveen and {Shapley}, Alice E. and {Erb}, Dawn K. and {Law}, David R.},
        title = "{The Gaseous Environment of High-z Galaxies: Precision Measurements of Neutral Hydrogen in the Circumgalactic Medium of z \raisebox{-0.5ex}\textasciitilde 2-3 Galaxies in the Keck Baryonic Structure Survey}",
      journal = {\apj},
         year = 2012,
        month = may,
       volume = {750},
       number = {1},
          eid = {67},
        pages = {67},
          doi = {10.1088/0004-637X/750/1/67},
archivePrefix = {arXiv},
       eprint = {1202.6055},
 primaryClass = {astro-ph.CO},
       adsurl = {https://ui.adsabs.harvard.edu/abs/2012ApJ...750...67R}
}

@ARTICLE{stei2014,
       author = {{Steidel}, Charles C. and {Rudie}, Gwen C. and {Strom}, Allison L. and {Pettini}, Max and {Reddy}, Naveen A. and {Shapley}, Alice E. and {Trainor}, Ryan F. and {Erb}, Dawn K. and {Turner}, Monica L. and {Konidaris}, Nicholas P. and {Kulas}, Kristin R. and {Mace}, Gregory and {Matthews}, Keith and {McLean}, Ian S.},
        title = "{Strong Nebular Line Ratios in the Spectra of z \raisebox{-0.5ex}\textasciitilde 2-3 Star Forming Galaxies: First Results from KBSS-MOSFIRE}",
      journal = {\apj},
         year = 2014,
        month = nov,
       volume = {795},
       number = {2},
          eid = {165},
        pages = {165},
          doi = {10.1088/0004-637X/795/2/165},
archivePrefix = {arXiv},
       eprint = {1405.5473},
 primaryClass = {astro-ph.GA},
       adsurl = {https://ui.adsabs.harvard.edu/abs/2014ApJ...795..165S}
}

@ARTICLE{stei2016,
       author = {{Steidel}, Charles C. and {Strom}, Allison L. and {Pettini}, Max and {Rudie}, Gwen C. and {Reddy}, Naveen A. and {Trainor}, Ryan F.},
        title = "{Reconciling the Stellar and Nebular Spectra of High-redshift Galaxies}",
      journal = {\apj},
         year = 2016,
        month = aug,
       volume = {826},
       number = {2},
          eid = {159},
        pages = {159},
          doi = {10.3847/0004-637X/826/2/159},
archivePrefix = {arXiv},
       eprint = {1605.07186},
 primaryClass = {astro-ph.GA},
       adsurl = {https://ui.adsabs.harvard.edu/abs/2016ApJ...826..159S}
}

@ARTICLE{stro2017,
       author = {{Strom}, Allison L. and {Steidel}, Charles C. and {Rudie}, Gwen C. and {Trainor}, Ryan F. and {Pettini}, Max and {Reddy}, Naveen A.},
        title = "{Nebular Emission Line Ratios in z ≃ 2-3 Star-forming Galaxies with KBSS-MOSFIRE: Exploring the Impact of Ionization, Excitation, and Nitrogen-to-Oxygen Ratio}",
      journal = {\apj},
         year = 2017,
        month = feb,
       volume = {836},
       number = {2},
          eid = {164},
        pages = {164},
          doi = {10.3847/1538-4357/836/2/164},
archivePrefix = {arXiv},
       eprint = {1608.02587},
 primaryClass = {astro-ph.GA},
       adsurl = {https://ui.adsabs.harvard.edu/abs/2017ApJ...836..164S}
}

@ARTICLE{stro2018,
       author = {{Strom}, Allison L. and {Steidel}, Charles C. and {Rudie}, Gwen C. and {Trainor}, Ryan F. and {Pettini}, Max},
        title = "{Measuring the Physical Conditions in High-redshift Star-forming Galaxies: Insights from KBSS-MOSFIRE}",
      journal = {\apj},
         year = 2018,
        month = dec,
       volume = {868},
       number = {2},
          eid = {117},
        pages = {117},
          doi = {10.3847/1538-4357/aae1a5},
archivePrefix = {arXiv},
       eprint = {1711.08820},
 primaryClass = {astro-ph.GA},
       adsurl = {https://ui.adsabs.harvard.edu/abs/2018ApJ...868..117S}
}

@ARTICLE{stro2023,
       author = {{Strom}, Allison L. and {Rudie}, Gwen C. and {Trainor}, Ryan F. and {Brammer}, Gabriel B. and {Maseda}, Michael V. and {Raptis}, Menelaos and {Rogers}, Noah S.~J. and {Steidel}, Charles C. and {Chen}, Yuguang and {Law}, David R.},
        title = "{CECILIA: The Faint Emission Line Spectrum of z   2-3 Star-forming Galaxies}",
      journal = {\apjl},
         year = 2023,
        month = nov,
       volume = {958},
       number = {1},
          eid = {L11},
        pages = {L11},
          doi = {10.3847/2041-8213/ad07dc},
archivePrefix = {arXiv},
       eprint = {2308.13508},
 primaryClass = {astro-ph.GA},
       adsurl = {https://ui.adsabs.harvard.edu/abs/2023ApJ...958L..11S}
}

@ARTICLE{trai2015,
       author = {{Trainor}, Ryan F. and {Steidel}, Charles C. and {Strom}, Allison L. and {Rudie}, Gwen C.},
        title = "{The Spectroscopic Properties of Ly{\ensuremath{\alpha}}-Emitters at z {\ensuremath{\sim}}2.7: Escaping Gas and Photons from Faint Galaxies}",
      journal = {\apj},
         year = 2015,
        month = aug,
       volume = {809},
       number = {1},
          eid = {89},
        pages = {89},
          doi = {10.1088/0004-637X/809/1/89},
archivePrefix = {arXiv},
       eprint = {1506.08205},
 primaryClass = {astro-ph.GA},
       adsurl = {https://ui.adsabs.harvard.edu/abs/2015ApJ...809...89T}
}

@ARTICLE{trem2004,
       author = {{Tremonti}, Christy A. and {Heckman}, Timothy M. and {Kauffmann}, Guinevere and {Brinchmann}, Jarle and {Charlot}, St\'{e}phane and {White}, Simon D.~M. and {Seibert}, Mark and {Peng}, Eric W. and {Schlegel}, David J. and {Uomoto}, Alan and {Fukugita}, Masataka and {Brinkmann}, Jon},
        title = "{The Origin of the Mass-Metallicity Relation: Insights from 53,000 Star-forming Galaxies in the Sloan Digital Sky Survey}",
      journal = {\apj},
         year = 2004,
        month = oct,
       volume = {613},
       number = {2},
        pages = {898-913},
          doi = {10.1086/423264},
archivePrefix = {arXiv},
       eprint = {astro-ph/0405537},
 primaryClass = {astro-ph},
       adsurl = {https://ui.adsabs.harvard.edu/abs/2004ApJ...613..898T}
}

@ARTICLE{Trainor2016,
       author = {{Trainor}, Ryan F. and {Strom}, Allison L. and {Steidel}, Charles C. and {Rudie}, Gwen C.},
        title = "{The Rest-frame Optical Spectroscopic Properties of Lyalpha-emitters at Z\raisebox{-0.5ex}\textasciitilde2.5: The Physical Origins of Strong Lyalpha Emission}",
      journal = {\apj},
         year = 2016,
        month = dec,
       volume = {832},
       number = {2},
          eid = {171},
        pages = {171},
          doi = {10.3847/0004-637X/832/2/171},
archivePrefix = {arXiv},
       eprint = {1608.07280},
 primaryClass = {astro-ph.GA},
       adsurl = {https://ui.adsabs.harvard.edu/abs/2016ApJ...832..171T}
}

@ARTICLE{Sanders2023,
       author = {{Sanders}, Ryan L. and {Shapley}, Alice E. and {Topping}, Michael W. and {Reddy}, Naveen A. and {Brammer}, Gabriel B.},
        title = "{Excitation and Ionization Properties of Star-forming Galaxies at z = 2.0-9.3 with JWST/NIRSpec}",
      journal = {\apj},
         year = 2023,
        month = sep,
       volume = {955},
       number = {1},
          eid = {54},
        pages = {54},
          doi = {10.3847/1538-4357/acedad},
archivePrefix = {arXiv},
       eprint = {2301.06696},
 primaryClass = {astro-ph.GA},
       adsurl = {https://ui.adsabs.harvard.edu/abs/2023ApJ...955...54S}
}

@ARTICLE{Kauffmann2003,
       author = {{Kauffmann}, Guinevere and {Heckman}, Timothy M. and {White}, Simon D.~M. and {Charlot}, St{\'e}phane and {Tremonti}, Christy and {Brinchmann}, Jarle and {Bruzual}, Gustavo and {Peng}, Eric W. and {Seibert}, Mark and {Bernardi}, Mariangela and {Blanton}, Michael and {Brinkmann}, Jon and {Castander}, Francisco and {Cs{\'a}bai}, Istvan and {Fukugita}, Masataka and {Ivezic}, Zeljko and {Munn}, Jeffrey A. and {Nichol}, Robert C. and {Padmanabhan}, Nikhil and {Thakar}, Aniruddha R. and {Weinberg}, David H. and {York}, Donald},
        title = "{Stellar masses and star formation histories for {}10$^{5}$ galaxies from the Sloan Digital Sky Survey}",
      journal = {\mnras},
         year = 2003,
        month = may,
       volume = {341},
       number = {1},
        pages = {33-53},
          doi = {10.1046/j.1365-8711.2003.06291.x},
archivePrefix = {arXiv},
       eprint = {astro-ph/0204055},
 primaryClass = {astro-ph},
       adsurl = {https://ui.adsabs.harvard.edu/abs/2003MNRAS.341...33K}
}

@ARTICLE{Baldwin1981,
       author = {{Baldwin}, J.~A. and {Phillips}, M.~M. and {Terlevich}, R.},
        title = "{Classification parameters for the emission-line spectra of extragalactic objects.}",
      journal = {\pasp},
         year = 1981,
        month = feb,
       volume = {93},
        pages = {5-19},
          doi = {10.1086/130766},
       adsurl = {https://ui.adsabs.harvard.edu/abs/1981PASP...93....5B}
}

@ARTICLE{Trainor2019,
       author = {{Trainor}, Ryan F. and {Strom}, Allison L. and {Steidel}, Charles C. and {Rudie}, Gwen C. and {Chen}, Yuguang and {Theios}, Rachel L.},
        title = "{Predicting Ly{\ensuremath{\alpha}} Emission from Galaxies via Empirical Markers of Production and Escape in the KBSS}",
      journal = {\apj},
         year = 2019,
        month = dec,
       volume = {887},
       number = {1},
          eid = {85},
        pages = {85},
          doi = {10.3847/1538-4357/ab4993},
archivePrefix = {arXiv},
       eprint = {1908.04794},
 primaryClass = {astro-ph.GA},
       adsurl = {https://ui.adsabs.harvard.edu/abs/2019ApJ...887...85T}
}

@ARTICLE{Shapley2019,
       author = {{Shapley}, Alice E. and {Sanders}, Ryan L. and {Shao}, Peng and {Reddy}, Naveen A. and {Kriek}, Mariska and {Coil}, Alison L. and {Mobasher}, Bahram and {Siana}, Brian and {Shivaei}, Irene and {Freeman}, William R. and {Azadi}, Mojegan and {Price}, Sedona H. and {Leung}, Gene C.~K. and {Fetherolf}, Tara and {de Groot}, Laura and {Zick}, Tom and {Fornasini}, Francesca M. and {Barro}, Guillermo},
        title = "{The MOSDEF Survey: Sulfur Emission-line Ratios Provide New Insights into Evolving Interstellar Medium Conditions at High Redshift}",
      journal = {\apjl},
         year = 2019,
        month = aug,
       volume = {881},
       number = {2},
          eid = {L35},
        pages = {L35},
          doi = {10.3847/2041-8213/ab385a},
archivePrefix = {arXiv},
       eprint = {1907.07189},
 primaryClass = {astro-ph.GA},
       adsurl = {https://ui.adsabs.harvard.edu/abs/2019ApJ...881L..35S}
}

@ARTICLE{Kennicutt1998,
       author = {{Kennicutt}, Robert C., Jr.},
        title = "{Star Formation in Galaxies Along the Hubble Sequence}",
      journal = {\araa},
         year = 1998,
        month = jan,
       volume = {36},
        pages = {189-232},
          doi = {10.1146/annurev.astro.36.1.189},
archivePrefix = {arXiv},
       eprint = {astro-ph/9807187},
 primaryClass = {astro-ph},
       adsurl = {https://ui.adsabs.harvard.edu/abs/1998ARA&A..36..189K}
}

@ARTICLE{Bian2018,
       author = {{Bian}, Fuyan and {Kewley}, Lisa J. and {Dopita}, Michael A.},
        title = "{{\textquotedblleft}Direct{\textquotedblright} Gas-phase Metallicity in Local Analogs of High-redshift Galaxies: Empirical Metallicity Calibrations for High-redshift Star-forming Galaxies}",
      journal = {\apj},
         year = 2018,
        month = jun,
       volume = {859},
       number = {2},
          eid = {175},
        pages = {175},
          doi = {10.3847/1538-4357/aabd74},
archivePrefix = {arXiv},
       eprint = {1805.08224},
 primaryClass = {astro-ph.GA},
       adsurl = {https://ui.adsabs.harvard.edu/abs/2018ApJ...859..175B}
}

@ARTICLE{Andrews2013,
       author = {{Andrews}, Brett H. and {Martini}, Paul},
        title = "{The Mass-Metallicity Relation with the Direct Method on Stacked Spectra of SDSS Galaxies}",
      journal = {\apj},
         year = 2013,
        month = mar,
       volume = {765},
       number = {2},
          eid = {140},
        pages = {140},
          doi = {10.1088/0004-637X/765/2/140},
archivePrefix = {arXiv},
       eprint = {1211.3418},
 primaryClass = {astro-ph.CO},
       adsurl = {https://ui.adsabs.harvard.edu/abs/2013ApJ...765..140A}
}

@ARTICLE{Horne1986,
       author = {{Horne}, K.},
        title = "{An optimal extraction algorithm for CCD spectroscopy.}",
      journal = {\pasp},
         year = 1986,
        month = jun,
       volume = {98},
        pages = {609-617},
          doi = {10.1086/131801},
       adsurl = {https://ui.adsabs.harvard.edu/abs/1986PASP...98..609H}
}

@ARTICLE{Sanders2024,
       author = {{Sanders}, Ryan L. and {Shapley}, Alice E. and {Topping}, Michael W. and {Reddy}, Naveen A. and {Brammer}, Gabriel B.},
        title = "{Direct T $_{e}$-based Metallicities of z = 2{\textendash}9 Galaxies with JWST/NIRSpec: Empirical Metallicity Calibrations Applicable from Reionization to Cosmic Noon}",
      journal = {\apj},
         year = 2024,
        month = feb,
       volume = {962},
       number = {1},
          eid = {24},
        pages = {24},
          doi = {10.3847/1538-4357/ad15fc},
archivePrefix = {arXiv},
       eprint = {2303.08149},
 primaryClass = {astro-ph.GA},
       adsurl = {https://ui.adsabs.harvard.edu/abs/2024ApJ...962...24S}
}

@ARTICLE{Nakajima22,
       author = {{Nakajima}, Kimihiko and {Ouchi}, Masami and {Xu}, Yi and {Rauch}, Michael and {Harikane}, Yuichi and {Nishigaki}, Moka and {Isobe}, Yuki and {Kusakabe}, Haruka and {Nagao}, Tohru and {Ono}, Yoshiaki and {Onodera}, Masato and {Sugahara}, Yuma and {Kim}, Ji Hoon and {Komiyama}, Yutaka and {Lee}, Chien-Hsiu and {Zahedy}, Fakhri S.},
        title = "{EMPRESS. V. Metallicity Diagnostics of Galaxies over 12 + log(O/H) ≃ 6.9-8.9 Established by a Local Galaxy Census: Preparing for JWST Spectroscopy}",
      journal = {\apjs},
         year = 2022,
        month = sep,
       volume = {262},
       number = {1},
          eid = {3},
        pages = {3},
          doi = {10.3847/1538-4365/ac7710},
archivePrefix = {arXiv},
       eprint = {2206.02824},
 primaryClass = {astro-ph.GA},
       adsurl = {https://ui.adsabs.harvard.edu/abs/2022ApJS..262....3N}
}

@ARTICLE{korh2025,
       author = {{Korhonen Cuestas}, Nathalie A. and {Strom}, Allison L. and {Miller}, Tim B. and {Steidel}, Charles C. and {Trainor}, Ryan F. and {Rudie}, Gwen C. and {Nu{\~n}ez}, Evan Haze},
        title = "{Exploring the Relationship Between Stellar Mass, Metallicity, and Star Formation Rate at $z \sim 2.3$ in KBSS-MOSFIRE}",
      journal = {arXiv e-prints},
         year = 2025,
        month = mar,
          eid = {arXiv:2503.10800},
        pages = {arXiv:2503.10800},
          doi = {10.48550/arXiv.2503.10800},
archivePrefix = {arXiv},
       eprint = {2503.10800},
 primaryClass = {astro-ph.GA},
       adsurl = {https://ui.adsabs.harvard.edu/abs/2025arXiv250310800K}
}

@ARTICLE{steidel2004,
       author = {{Steidel}, Charles C. and {Shapley}, Alice E. and {Pettini}, Max and {Adelberger}, Kurt L. and {Erb}, Dawn K. and {Reddy}, Naveen A. and {Hunt}, Matthew P.},
        title = "{A Survey of Star-forming Galaxies in the 1.4<\raisebox{-0.5ex}\textasciitildeZ<\raisebox{-0.5ex}\textasciitilde 2.5 Redshift Desert: Overview}",
      journal = {\apj},
         year = 2004,
        month = apr,
       volume = {604},
       number = {2},
        pages = {534-550},
          doi = {10.1086/381960},
archivePrefix = {arXiv},
       eprint = {astro-ph/0401439},
 primaryClass = {astro-ph},
       adsurl = {https://ui.adsabs.harvard.edu/abs/2004ApJ...604..534S}
}

@ARTICLE{oke1995,
       author = {{Oke}, J.~B. and {Cohen}, J.~G. and {Carr}, M. and {Cromer}, J. and {Dingizian}, A. and {Harris}, F.~H. and {Labrecque}, S. and {Lucinio}, R. and {Schaal}, W. and {Epps}, H. and {Miller}, J.},
        title = "{The Keck Low-Resolution Imaging Spectrometer}",
      journal = {\pasp},
         year = 1995,
        month = apr,
       volume = {107},
        pages = {375},
          doi = {10.1086/133562},
       adsurl = {https://ui.adsabs.harvard.edu/abs/1995PASP..107..375O}
}

@INPROCEEDINGS{mclean2012,
       author = {{McLean}, Ian S. and {Steidel}, Charles C. and {Epps}, Harland W. and {Konidaris}, Nicholas and {Matthews}, Keith Y. and {Adkins}, Sean and {Aliado}, Theodore and {Brims}, George and {Canfield}, John M. and {Cromer}, John L. and {Fucik}, Jason and {Kulas}, Kristin and {Mace}, Greg and {Magnone}, Ken and {Rodriguez}, Hector and {Rudie}, Gwen and {Trainor}, Ryan and {Wang}, Eric and {Weber}, Bob and {Weiss}, Jason},
        title = "{MOSFIRE, the multi-object spectrometer for infra-red exploration at the Keck Observatory}",
    booktitle = {Ground-based and Airborne Instrumentation for Astronomy IV},
         year = 2012,
       editor = {{McLean}, Ian S. and {Ramsay}, Suzanne K. and {Takami}, Hideki},
       series = {Society of Photo-Optical Instrumentation Engineers (SPIE) Conference Series},
       volume = {8446},
        month = sep,
          eid = {84460J},
        pages = {84460J},
          doi = {10.1117/12.924794},
       adsurl = {https://ui.adsabs.harvard.edu/abs/2012SPIE.8446E..0JM}
}

@ARTICLE{Stanway2018,
       author = {{Stanway}, E.~R. and {Eldridge}, J.~J.},
        title = "{Re-evaluating old stellar populations}",
      journal = {\mnras},
         year = 2018,
        month = sep,
       volume = {479},
       number = {1},
        pages = {75-93},
          doi = {10.1093/mnras/sty1353},
archivePrefix = {arXiv},
       eprint = {1805.08784},
 primaryClass = {astro-ph.GA},
       adsurl = {https://ui.adsabs.harvard.edu/abs/2018MNRAS.479...75S}
}

@ARTICLE{Chabrier2003,
       author = {{Chabrier}, Gilles},
        title = "{Galactic Stellar and Substellar Initial Mass Function}",
      journal = {\pasp},
         year = 2003,
        month = jul,
       volume = {115},
       number = {809},
        pages = {763-795},
          doi = {10.1086/376392},
archivePrefix = {arXiv},
       eprint = {astro-ph/0304382},
 primaryClass = {astro-ph},
       adsurl = {https://ui.adsabs.harvard.edu/abs/2003PASP..115..763C}
}

@ARTICLE{Sanders2021,
       author = {{Sanders}, Ryan L. and {Shapley}, Alice E. and {Jones}, Tucker and {Reddy}, Naveen A. and {Kriek}, Mariska and {Siana}, Brian and {Coil}, Alison L. and {Mobasher}, Bahram and {Shivaei}, Irene and {Dav{\'e}}, Romeel and {Azadi}, Mojegan and {Price}, Sedona H. and {Leung}, Gene and {Freeman}, William R. and {Fetherolf}, Tara and {de Groot}, Laura and {Zick}, Tom and {Barro}, Guillermo},
        title = "{The MOSDEF Survey: The Evolution of the Mass-Metallicity Relation from z = 0 to z 3.3}",
      journal = {\apj},
         year = 2021,
        month = jun,
       volume = {914},
       number = {1},
          eid = {19},
        pages = {19},
          doi = {10.3847/1538-4357/abf4c1},
archivePrefix = {arXiv},
       eprint = {2009.07292},
 primaryClass = {astro-ph.GA},
       adsurl = {https://ui.adsabs.harvard.edu/abs/2021ApJ...914...19S}
}

@ARTICLE{2019A&ARv..27....3M,
       author = {{Maiolino}, R. and {Mannucci}, F.},
        title = "{De re metallica: the cosmic chemical evolution of galaxies}",
      journal = {\aapr},
         year = 2019,
        month = feb,
       volume = {27},
       number = {1},
          eid = {3},
        pages = {3},
          doi = {10.1007/s00159-018-0112-2},
archivePrefix = {arXiv},
       eprint = {1811.09642},
 primaryClass = {astro-ph.GA},
       adsurl = {https://ui.adsabs.harvard.edu/abs/2019A&ARv..27....3M}
}

@ARTICLE{York2000,
       author = {{York}, Donald G. and {Adelman}, J. and {Anderson}, Jr., John E. and {Anderson}, Scott F. and {Annis}, James and {Bahcall}, Neta A. and {Bakken}, J.~A. and {Barkhouser}, Robert and {Bastian}, Steven and {Berman}, Eileen and {Boroski}, William N. and {Bracker}, Steve and {Briegel}, Charlie and {Briggs}, John W. and {Brinkmann}, J. and {Brunner}, Robert and {Burles}, Scott and {Carey}, Larry and {Carr}, Michael A. and {Castander}, Francisco J. and {Chen}, Bing and {Colestock}, Patrick L. and {Connolly}, A.~J. and {Crocker}, J.~H. and {Csabai}, Istv{\'a}n and {Czarapata}, Paul C. and {Davis}, John Eric and {Doi}, Mamoru and {Dombeck}, Tom and {Eisenstein}, Daniel and {Ellman}, Nancy and {Elms}, Brian R. and {Evans}, Michael L. and {Fan}, Xiaohui and {Federwitz}, Glenn R. and {Fiscelli}, Larry and {Friedman}, Scott and {Frieman}, Joshua A. and {Fukugita}, Masataka and {Gillespie}, Bruce and {Gunn}, James E. and {Gurbani}, Vijay K. and {de Haas}, Ernst and {Haldeman}, Merle and {Harris}, Frederick H. and {Hayes}, J. and {Heckman}, Timothy M. and {Hennessy}, G.~S. and {Hindsley}, Robert B. and {Holm}, Scott and {Holmgren}, Donald J. and {Huang}, Chi-hao and {Hull}, Charles and {Husby}, Don and {Ichikawa}, Shin-Ichi and {Ichikawa}, Takashi and {Ivezi{\'c}}, {\v{Z}}eljko and {Kent}, Stephen and {Kim}, Rita S.~J. and {Kinney}, E. and {Klaene}, Mark and {Kleinman}, A.~N. and {Kleinman}, S. and {Knapp}, G.~R. and {Korienek}, John and {Kron}, Richard G. and {Kunszt}, Peter Z. and {Lamb}, D.~Q. and {Lee}, B. and {Leger}, R. French and {Limmongkol}, Siriluk and {Lindenmeyer}, Carl and {Long}, Daniel C. and {Loomis}, Craig and {Loveday}, Jon and {Lucinio}, Rich and {Lupton}, Robert H. and {MacKinnon}, Bryan and {Mannery}, Edward J. and {Mantsch}, P.~M. and {Margon}, Bruce and {McGehee}, Peregrine and {McKay}, Timothy A. and {Meiksin}, Avery and {Merelli}, Aronne and {Monet}, David G. and {Munn}, Jeffrey A. and {Narayanan}, Vijay K. and {Nash}, Thomas and {Neilsen}, Eric and {Neswold}, Rich and {Newberg}, Heidi Jo and {Nichol}, R.~C. and {Nicinski}, Tom and {Nonino}, Mario and {Okada}, Norio and {Okamura}, Sadanori and {Ostriker}, Jeremiah P. and {Owen}, Russell and {Pauls}, A. George and {Peoples}, John and {Peterson}, R.~L. and {Petravick}, Donald and {Pier}, Jeffrey R. and {Pope}, Adrian and {Pordes}, Ruth and {Prosapio}, Angela and {Rechenmacher}, Ron and {Quinn}, Thomas R. and {Richards}, Gordon T. and {Richmond}, Michael W. and {Rivetta}, Claudio H. and {Rockosi}, Constance M. and {Ruthmansdorfer}, Kurt and {Sandford}, Dale and {Schlegel}, David J. and {Schneider}, Donald P. and {Sekiguchi}, Maki and {Sergey}, Gary and {Shimasaku}, Kazuhiro and {Siegmund}, Walter A. and {Smee}, Stephen and {Smith}, J. Allyn and {Snedden}, S. and {Stone}, R. and {Stoughton}, Chris and {Strauss}, Michael A. and {Stubbs}, Christopher and {SubbaRao}, Mark and {Szalay}, Alexander S. and {Szapudi}, Istvan and {Szokoly}, Gyula P. and {Thakar}, Anirudda R. and {Tremonti}, Christy and {Tucker}, Douglas L. and {Uomoto}, Alan and {Vanden Berk}, Dan and {Vogeley}, Michael S. and {Waddell}, Patrick and {Wang}, Shu-i. and {Watanabe}, Masaru and {Weinberg}, David H. and {Yanny}, Brian and {Yasuda}, Naoki and {SDSS Collaboration}},
        title = "{The Sloan Digital Sky Survey: Technical Summary}",
      journal = {\aj},
         year = 2000,
        month = sep,
       volume = {120},
       number = {3},
        pages = {1579-1587},
          doi = {10.1086/301513},
archivePrefix = {arXiv},
       eprint = {astro-ph/0006396},
 primaryClass = {astro-ph},
       adsurl = {https://ui.adsabs.harvard.edu/abs/2000AJ....120.1579Y}
}

@ARTICLE{Abazajian2003,
       author = {{Abazajian}, Kevork and {Adelman-McCarthy}, Jennifer K. and {Ag{\"u}eros}, Marcel A. and {Allam}, Sahar S. and {Anderson}, Scott F. and {Annis}, James and {Bahcall}, Neta A. and {Baldry}, Ivan K. and {Bastian}, Steven and {Berlind}, Andreas and {Bernardi}, Mariangela and {Blanton}, Michael R. and {Blythe}, Norman and {Bochanski}, Jr., John J. and {Boroski}, William N. and {Brewington}, Howard and {Briggs}, John W. and {Brinkmann}, J. and {Brunner}, Robert J. and {Budav{\'a}ri}, Tam{\'a}s and {Carey}, Larry N. and {Carr}, Michael A. and {Castander}, Francisco J. and {Chiu}, Kuenley and {Collinge}, Matthew J. and {Connolly}, A.~J. and {Covey}, Kevin R. and {Csabai}, Istv{\'a}n and {Dalcanton}, Julianne J. and {Dodelson}, Scott and {Doi}, Mamoru and {Dong}, Feng and {Eisenstein}, Daniel J. and {Evans}, Michael L. and {Fan}, Xiaohui and {Feldman}, Paul D. and {Finkbeiner}, Douglas P. and {Friedman}, Scott D. and {Frieman}, Joshua A. and {Fukugita}, Masataka and {Gal}, Roy R. and {Gillespie}, Bruce and {Glazebrook}, Karl and {Gonzalez}, Carlos F. and {Gray}, Jim and {Grebel}, Eva K. and {Grodnicki}, Lauren and {Gunn}, James E. and {Gurbani}, Vijay K. and {Hall}, Patrick B. and {Hao}, Lei and {Harbeck}, Daniel and {Harris}, Frederick H. and {Harris}, Hugh C. and {Harvanek}, Michael and {Hawley}, Suzanne L. and {Heckman}, Timothy M. and {Helmboldt}, J.~F. and {Hendry}, John S. and {Hennessy}, Gregory S. and {Hindsley}, Robert B. and {Hogg}, David W. and {Holmgren}, Donald J. and {Holtzman}, Jon A. and {Homer}, Lee and {Hui}, Lam and {Ichikawa}, Shin-ichi and {Ichikawa}, Takashi and {Inkmann}, John P. and {Ivezi{\'c}}, {\v{Z}}eljko and {Jester}, Sebastian and {Johnston}, David E. and {Jordan}, Beatrice and {Jordan}, Wendell P. and {Jorgensen}, Anders M. and {Juri{\'c}}, Mario and {Kauffmann}, Guinevere and {Kent}, Stephen M. and {Kleinman}, S.~J. and {Knapp}, G.~R. and {Kniazev}, Alexei Y. and {Kron}, Richard G. and {Krzesi{\'n}ski}, Jurek and {Kunszt}, Peter Z. and {Kuropatkin}, Nickolai and {Lamb}, Donald Q. and {Lampeitl}, Hubert and {Laubscher}, Bryan E. and {Lee}, Brian C. and {Leger}, R. French and {Li}, Nolan and {Lidz}, Adam and {Lin}, Huan and {Loh}, Yeong-Shang and {Long}, Daniel C. and {Loveday}, Jon and {Lupton}, Robert H. and {Malik}, Tanu and {Margon}, Bruce and {McGehee}, Peregrine M. and {McKay}, Timothy A. and {Meiksin}, Avery and {Miknaitis}, Gajus A. and {Moorthy}, Bhasker K. and {Munn}, Jeffrey A. and {Murphy}, Tara and {Nakajima}, Reiko and {Narayanan}, Vijay K. and {Nash}, Thomas and {Neilsen}, Jr., Eric H. and {Newberg}, Heidi Jo and {Newman}, Peter R. and {Nichol}, Robert C. and {Nicinski}, Tom and {Nieto-Santisteban}, Maria and {Nitta}, Atsuko and {Odenkirchen}, Michael and {Okamura}, Sadanori and {Ostriker}, Jeremiah P. and {Owen}, Russell and {Padmanabhan}, Nikhil and {Peoples}, John and {Pier}, Jeffrey R. and {Pindor}, Bartosz and {Pope}, Adrian C. and {Quinn}, Thomas R. and {Rafikov}, R.~R. and {Raymond}, Sean N. and {Richards}, Gordon T. and {Richmond}, Michael W. and {Rix}, Hans-Walter and {Rockosi}, Constance M. and {Schaye}, Joop and {Schlegel}, David J. and {Schneider}, Donald P. and {Schroeder}, Joshua and {Scranton}, Ryan and {Sekiguchi}, Maki and {Seljak}, Uro{\v{s}} and {Sergey}, Gary and {Sesar}, Branimir and {Sheldon}, Erin and {Shimasaku}, Kazu and {Siegmund}, Walter A. and {Silvestri}, Nicole M. and {Sinisgalli}, Allan J. and {Sirko}, Edwin and {Smith}, J. Allyn and {Smol{\v{c}}i{\'c}}, Vernesa and {Snedden}, Stephanie A. and {Stebbins}, Albert and {Steinhardt}, Charles and {Stinson}, Gregory and {Stoughton}, Chris and {Strateva}, Iskra V. and {Strauss}, Michael A. and {SubbaRao}, Mark and {Szalay}, Alexander S. and {Szapudi}, Istv{\'a}n and {Szkody}, Paula and {Tasca}, Lidia and {Tegmark}, Max and {Thakar}, Aniruddha R. and {Tremonti}, Christy and {Tucker}, Douglas L. and {Uomoto}, Alan and {Vanden Berk}, Daniel E. and {Vandenberg}, Jan and {Vogeley}, Michael S. and {Voges}, Wolfgang and {Vogt}, Nicole P. and {Walkowicz}, Lucianne M. and {Weinberg}, David H. and {West}, Andrew A. and {White}, Simon D.~M. and {Wilhite}, Brian C. and {Willman}, Beth and {Xu}, Yongzhong and {Yanny}, Brian and {Yarger}, Jean and {Yasuda}, Naoki and {Yip}, Ching-Wa and {Yocum}, D.~R. and {York}, Donald G. and {Zakamska}, Nadia L. and {Zehavi}, Idit and {Zheng}, Wei and {Zibetti}, Stefano and {Zucker}, Daniel B.},
        title = "{The First Data Release of the Sloan Digital Sky Survey}",
      journal = {\aj},
         year = 2003,
        month = oct,
       volume = {126},
       number = {4},
        pages = {2081-2086},
          doi = {10.1086/378165},
archivePrefix = {arXiv},
       eprint = {astro-ph/0305492},
 primaryClass = {astro-ph},
       adsurl = {https://ui.adsabs.harvard.edu/abs/2003AJ....126.2081A}
}

@ARTICLE{Kirby2013,
       author = {{Kirby}, Evan N. and {Cohen}, Judith G. and {Guhathakurta}, Puragra and {Cheng}, Lucy and {Bullock}, James S. and {Gallazzi}, Anna},
        title = "{The Universal Stellar Mass-Stellar Metallicity Relation for Dwarf Galaxies}",
      journal = {\apj},
         year = 2013,
        month = dec,
       volume = {779},
       number = {2},
          eid = {102},
        pages = {102},
          doi = {10.1088/0004-637X/779/2/102},
archivePrefix = {arXiv},
       eprint = {1310.0814},
 primaryClass = {astro-ph.GA},
       adsurl = {https://ui.adsabs.harvard.edu/abs/2013ApJ...779..102K}
}

@ARTICLE{Kewley2008,
       author = {{Kewley}, Lisa J. and {Ellison}, Sara L.},
        title = "{Metallicity Calibrations and the Mass-Metallicity Relation for Star-forming Galaxies}",
      journal = {\apj},
         year = 2008,
        month = jul,
       volume = {681},
       number = {2},
        pages = {1183-1204},
          doi = {10.1086/587500},
archivePrefix = {arXiv},
       eprint = {0801.1849},
 primaryClass = {astro-ph},
       adsurl = {https://ui.adsabs.harvard.edu/abs/2008ApJ...681.1183K}
}

@ARTICLE{Lequeux1979,
       author = {{Lequeux}, J. and {Peimbert}, M. and {Rayo}, J.~F. and {Serrano}, A. and {Torres-Peimbert}, S.},
        title = "{Chemical Composition and Evolution of Irregular and Blue Compact Galaxies}",
      journal = {\aap},
         year = 1979,
        month = dec,
       volume = {80},
        pages = {155},
       adsurl = {https://ui.adsabs.harvard.edu/abs/1979A&A....80..155L}
}

@ARTICLE{Li2023,
       author = {{Li}, Mingyu and {Cai}, Zheng and {Bian}, Fuyan and {Lin}, Xiaojing and {Li}, Zihao and {Wu}, Yunjing and {Sun}, Fengwu and {Zhang}, Shiwu and {Golden-Marx}, Emmet and {Sun}, Zechang and {Zou}, Siwei and {Fan}, Xiaohui and {Egami}, Eiichi and {Charlot}, Stephane and {Bruzual}, Gustavo and {Chevallard}, Jacopo},
        title = "{The Mass-Metallicity Relation of Dwarf Galaxies at Cosmic Noon from JWST Observations}",
      journal = {\apjl},
         year = 2023,
        month = sep,
       volume = {955},
       number = {1},
          eid = {L18},
        pages = {L18},
          doi = {10.3847/2041-8213/acf470},
archivePrefix = {arXiv},
       eprint = {2211.01382},
 primaryClass = {astro-ph.GA},
       adsurl = {https://ui.adsabs.harvard.edu/abs/2023ApJ...955L..18L}
}

@ARTICLE{Curti2024,
       author = {{Curti}, Mirko and {Maiolino}, Roberto and {Curtis-Lake}, Emma and {Chevallard}, Jacopo and {Carniani}, Stefano and {D'Eugenio}, Francesco and {Looser}, Tobias J. and {Scholtz}, Jan and {Charlot}, Stephane and {Cameron}, Alex and {{\"U}bler}, Hannah and {Witstok}, Joris and {Boyett}, Kristian and {Laseter}, Isaac and {Sandles}, Lester and {Arribas}, Santiago and {Bunker}, Andrew and {Giardino}, Giovanna and {Maseda}, Michael V. and {Rawle}, Tim and {Rodr{\'\i}guez Del Pino}, Bruno and {Smit}, Renske and {Willott}, Chris J. and {Eisenstein}, Daniel J. and {Hausen}, Ryan and {Johnson}, Benjamin and {Rieke}, Marcia and {Robertson}, Brant and {Tacchella}, Sandro and {Williams}, Christina C. and {Willmer}, Christopher and {Baker}, William M. and {Bhatawdekar}, Rachana and {Egami}, Eiichi and {Helton}, Jakob M. and {Ji}, Zhiyuan and {Kumari}, Nimisha and {Perna}, Michele and {Shivaei}, Irene and {Sun}, Fengwu},
        title = "{JADES: Insights into the low-mass end of the mass-metallicity-SFR relation at 3 < z < 10 from deep JWST/NIRSpec spectroscopy}",
      journal = {\aap},
         year = 2024,
        month = apr,
       volume = {684},
          eid = {A75},
        pages = {A75},
          doi = {10.1051/0004-6361/202346698},
archivePrefix = {arXiv},
       eprint = {2304.08516},
 primaryClass = {astro-ph.GA},
       adsurl = {https://ui.adsabs.harvard.edu/abs/2024A&A...684A..75C}
}

@ARTICLE{Maiolino2008,
       author = {{Maiolino}, R. and {Nagao}, T. and {Grazian}, A. and {Cocchia}, F. and {Marconi}, A. and {Mannucci}, F. and {Cimatti}, A. and {Pipino}, A. and {Ballero}, S. and {Calura}, F. and {Chiappini}, C. and {Fontana}, A. and {Granato}, G.~L. and {Matteucci}, F. and {Pastorini}, G. and {Pentericci}, L. and {Risaliti}, G. and {Salvati}, M. and {Silva}, L.},
        title = "{AMAZE. I. The evolution of the mass-metallicity relation at z > 3}",
      journal = {\aap},
         year = 2008,
        month = sep,
       volume = {488},
       number = {2},
        pages = {463-479},
          doi = {10.1051/0004-6361:200809678},
archivePrefix = {arXiv},
       eprint = {0806.2410},
 primaryClass = {astro-ph},
       adsurl = {https://ui.adsabs.harvard.edu/abs/2008A&A...488..463M}
}

@ARTICLE{Curti2020,
       author = {{Curti}, Mirko and {Mannucci}, Filippo and {Cresci}, Giovanni and {Maiolino}, Roberto},
        title = "{The mass-metallicity and the fundamental metallicity relation revisited on a fully T$_{e}$-based abundance scale for galaxies}",
      journal = {\mnras},
         year = 2020,
        month = jan,
       volume = {491},
       number = {1},
        pages = {944-964},
          doi = {10.1093/mnras/stz2910},
archivePrefix = {arXiv},
       eprint = {1910.00597},
 primaryClass = {astro-ph.GA},
       adsurl = {https://ui.adsabs.harvard.edu/abs/2020MNRAS.491..944C}
}

@ARTICLE{Kewley2002,
       author = {{Kewley}, L.~J. and {Dopita}, M.~A.},
        title = "{Using Strong Lines to Estimate Abundances in Extragalactic H II Regions and Starburst Galaxies}",
      journal = {\apjs},
         year = 2002,
        month = sep,
       volume = {142},
       number = {1},
        pages = {35-52},
          doi = {10.1086/341326},
archivePrefix = {arXiv},
       eprint = {astro-ph/0206495},
 primaryClass = {astro-ph},
       adsurl = {https://ui.adsabs.harvard.edu/abs/2002ApJS..142...35K}
}

@ARTICLE{Scholte2025,
       author = {{Scholte}, D. and {Cullen}, F. and {Carnall}, A.~C. and {Arellano-C{\'o}rdova}, K.~Z. and {Stanton}, T.~M. and {Barrufet}, L. and {Begley}, R. and {Bondestam}, C. and {Donnan}, C.~T. and {Dunlop}, J.~S. and {Leung}, H. -H. and {McLeod}, D.~J. and {McLure}, R.~J. and {Moustakas}, J.~M. and {Pollock}, C.~L. and {Shapley}, A.~E. and {Stevenson}, S. and {Zou}, H.},
        title = "{The JWST EXCELS survey: probing strong-line diagnostics and the chemical evolution of galaxies over cosmic time using T$_{e}$-metallicities}",
      journal = {\mnras},
         year = 2025,
        month = jun,
       volume = {540},
       number = {2},
        pages = {1800-1826},
          doi = {10.1093/mnras/staf834},
archivePrefix = {arXiv},
       eprint = {2502.10499},
 primaryClass = {astro-ph.GA},
       adsurl = {https://ui.adsabs.harvard.edu/abs/2025MNRAS.540.1800S}
}

@ARTICLE{Strom2022,
       author = {{Strom}, Allison L. and {Rudie}, Gwen C. and {Steidel}, Charles C. and {Trainor}, Ryan F.},
        title = "{Chemical Abundance Scaling Relations for Multiple Elements in z ≃ 2-3 Star-forming Galaxies}",
      journal = {\apj},
         year = 2022,
        month = feb,
       volume = {925},
       number = {2},
          eid = {116},
        pages = {116},
          doi = {10.3847/1538-4357/ac38a3},
archivePrefix = {arXiv},
       eprint = {2111.06410},
 primaryClass = {astro-ph.GA},
       adsurl = {https://ui.adsabs.harvard.edu/abs/2022ApJ...925..116S}
}

@ARTICLE{Lee2006,
       author = {{Lee}, Henry and {Skillman}, Evan D. and {Cannon}, John M. and {Jackson}, Dale C. and {Gehrz}, Robert D. and {Polomski}, Elisha F. and {Woodward}, Charles E.},
        title = "{On Extending the Mass-Metallicity Relation of Galaxies by 2.5 Decades in Stellar Mass}",
      journal = {\apj},
         year = 2006,
        month = aug,
       volume = {647},
       number = {2},
        pages = {970-983},
          doi = {10.1086/505573},
archivePrefix = {arXiv},
       eprint = {astro-ph/0605036},
 primaryClass = {astro-ph},
       adsurl = {https://ui.adsabs.harvard.edu/abs/2006ApJ...647..970L}
}

@ARTICLE{Ellison2008,
       author = {{Ellison}, Sara L. and {Patton}, David R. and {Simard}, Luc and {McConnachie}, Alan W.},
        title = "{Clues to the Origin of the Mass-Metallicity Relation: Dependence on Star Formation Rate and Galaxy Size}",
      journal = {\apjl},
         year = 2008,
        month = jan,
       volume = {672},
       number = {2},
        pages = {L107},
          doi = {10.1086/527296},
archivePrefix = {arXiv},
       eprint = {0711.4833},
 primaryClass = {astro-ph},
       adsurl = {https://ui.adsabs.harvard.edu/abs/2008ApJ...672L.107E}
}

@ARTICLE{Hughes2013,
       author = {{Hughes}, T.~M. and {Cortese}, L. and {Boselli}, A. and {Gavazzi}, G. and {Davies}, J.~I.},
        title = "{The role of cold gas and environment on the stellar mass-metallicity relation of nearby galaxies}",
      journal = {\aap},
         year = 2013,
        month = feb,
       volume = {550},
          eid = {A115},
        pages = {A115},
          doi = {10.1051/0004-6361/201218822},
archivePrefix = {arXiv},
       eprint = {1207.4191},
 primaryClass = {astro-ph.CO},
       adsurl = {https://ui.adsabs.harvard.edu/abs/2013A&A...550A.115H}
}

@ARTICLE{Henry2013,
       author = {{Henry}, Alaina and {Scarlata}, Claudia and {Dom{\'\i}nguez}, Alberto and {Malkan}, Matthew and {Martin}, Crystal L. and {Siana}, Brian and {Atek}, Hakim and {Bedregal}, Alejandro G. and {Colbert}, James W. and {Rafelski}, Marc and {Ross}, Nathaniel and {Teplitz}, Harry and {Bunker}, Andrew J. and {Dressler}, Alan and {Hathi}, Nimish and {Masters}, Daniel and {McCarthy}, Patrick and {Straughn}, Amber},
        title = "{Low Masses and High Redshifts: The Evolution of the Mass-Metallicity Relation}",
      journal = {\apjl},
         year = 2013,
        month = oct,
       volume = {776},
       number = {2},
          eid = {L27},
        pages = {L27},
          doi = {10.1088/2041-8205/776/2/L27},
archivePrefix = {arXiv},
       eprint = {1309.4458},
 primaryClass = {astro-ph.CO},
       adsurl = {https://ui.adsabs.harvard.edu/abs/2013ApJ...776L..27H}
}

@ARTICLE{Rogers2024,
       author = {{Rogers}, Noah S.~J. and {Strom}, Allison L. and {Rudie}, Gwen C. and {Trainor}, Ryan F. and {Raptis}, Menelaos and {von Raesfeld}, Caroline},
        title = "{CECILIA: Direct O, N, S, and Ar Abundances in Q2343-D40, a Galaxy at z {\ensuremath{\sim}} 3}",
      journal = {\apjl},
         year = 2024,
        month = mar,
       volume = {964},
       number = {1},
          eid = {L12},
        pages = {L12},
          doi = {10.3847/2041-8213/ad2f37},
archivePrefix = {arXiv},
       eprint = {2312.08427},
 primaryClass = {astro-ph.GA},
       adsurl = {https://ui.adsabs.harvard.edu/abs/2024ApJ...964L..12R}
}

@ARTICLE{Dekel1986,
       author = {{Dekel}, A. and {Silk}, J.},
        title = "{The Origin of Dwarf Galaxies, Cold Dark Matter, and Biased Galaxy Formation}",
      journal = {\apj},
         year = 1986,
        month = apr,
       volume = {303},
        pages = {39},
          doi = {10.1086/164050},
       adsurl = {https://ui.adsabs.harvard.edu/abs/1986ApJ...303...39D}
}

@ARTICLE{Mac_Low1999,
       author = {{Mac Low}, Mordecai-Mark and {Ferrara}, Andrea},
        title = "{Starburst-driven Mass Loss from Dwarf Galaxies: Efficiency and Metal Ejection}",
      journal = {\apj},
         year = 1999,
        month = mar,
       volume = {513},
       number = {1},
        pages = {142-155},
          doi = {10.1086/306832},
archivePrefix = {arXiv},
       eprint = {astro-ph/9801237},
 primaryClass = {astro-ph},
       adsurl = {https://ui.adsabs.harvard.edu/abs/1999ApJ...513..142M}
}

@ARTICLE{2014A&A...563A..58T,
       author = {{Troncoso}, P. and {Maiolino}, R. and {Sommariva}, V. and {Cresci}, G. and {Mannucci}, F. and {Marconi}, A. and {Meneghetti}, M. and {Grazian}, A. and {Cimatti}, A. and {Fontana}, A. and {Nagao}, T. and {Pentericci}, L.},
        title = "{Metallicity evolution, metallicity gradients, and gas fractions at z \raisebox{-0.5ex}\textasciitilde 3.4}",
      journal = {\aap},
         year = 2014,
        month = mar,
       volume = {563},
          eid = {A58},
        pages = {A58},
          doi = {10.1051/0004-6361/201322099},
archivePrefix = {arXiv},
       eprint = {1311.4576},
 primaryClass = {astro-ph.CO},
       adsurl = {https://ui.adsabs.harvard.edu/abs/2014A&A...563A..58T}
}

@ARTICLE{2015ApJ...799..209W,
       author = {{Wisnioski}, E. and {F{\"o}rster Schreiber}, N.~M. and {Wuyts}, S. and {Wuyts}, E. and {Bandara}, K. and {Wilman}, D. and {Genzel}, R. and {Bender}, R. and {Davies}, R. and {Fossati}, M. and {Lang}, P. and {Mendel}, J.~T. and {Beifiori}, A. and {Brammer}, G. and {Chan}, J. and {Fabricius}, M. and {Fudamoto}, Y. and {Kulkarni}, S. and {Kurk}, J. and {Lutz}, D. and {Nelson}, E.~J. and {Momcheva}, I. and {Rosario}, D. and {Saglia}, R. and {Seitz}, S. and {Tacconi}, L.~J. and {van Dokkum}, P.~G.},
        title = "{The KMOS$^{3D}$ Survey: Design, First Results, and the Evolution of Galaxy Kinematics from 0.7 <= z <= 2.7}",
      journal = {\apj},
         year = 2015,
        month = feb,
       volume = {799},
       number = {2},
          eid = {209},
        pages = {209},
          doi = {10.1088/0004-637X/799/2/209},
archivePrefix = {arXiv},
       eprint = {1409.6791},
 primaryClass = {astro-ph.GA},
       adsurl = {https://ui.adsabs.harvard.edu/abs/2015ApJ...799..209W}
}

@ARTICLE{Jones2015,
       author = {{Jones}, Tucker and {Martin}, Crystal and {Cooper}, Michael C.},
        title = "{Temperature-based Metallicity Measurements at z=0.8: Direct Calibration of Strong-line Diagnostics at Intermediate Redshift}",
      journal = {\apj},
         year = 2015,
        month = nov,
       volume = {813},
       number = {2},
          eid = {126},
        pages = {126},
          doi = {10.1088/0004-637X/813/2/126},
archivePrefix = {arXiv},
       eprint = {1504.02417},
 primaryClass = {astro-ph.GA},
       adsurl = {https://ui.adsabs.harvard.edu/abs/2015ApJ...813..126J}
}

@ARTICLE{Henry2021,
       author = {{Henry}, Alaina and {Rafelski}, Marc and {Sunnquist}, Ben and {Pirzkal}, Norbert and {Pacifici}, Camilla and {Atek}, Hakim and {Bagley}, Micaela and {Baronchelli}, Ivano and {Barro}, Guillermo and {Bunker}, Andrew J. and {Colbert}, James and {Dai}, Y. Sophia and {Elmegreen}, Bruce G. and {Elmegreen}, Debra Meloy and {Finkelstein}, Steven and {Kocevski}, Dale and {Koekemoer}, Anton and {Malkan}, Matthew and {Martin}, Crystal L. and {Mehta}, Vihang and {Pahl}, Anthony and {Papovich}, Casey and {Rutkowski}, Michael and {S{\'a}nchez Almeida}, Jorge and {Scarlata}, Claudia and {Snyder}, Gregory and {Teplitz}, Harry},
        title = "{The Mass-Metallicity Relation at z   1-2 and Its Dependence on the Star Formation Rate}",
      journal = {\apj},
         year = 2021,
        month = oct,
       volume = {919},
       number = {2},
          eid = {143},
        pages = {143},
          doi = {10.3847/1538-4357/ac1105},
archivePrefix = {arXiv},
       eprint = {2107.00672},
 primaryClass = {astro-ph.GA},
       adsurl = {https://ui.adsabs.harvard.edu/abs/2021ApJ...919..143H}
}

@ARTICLE{Mannucci2010,
       author = {{Mannucci}, F. and {Cresci}, G. and {Maiolino}, R. and {Marconi}, A. and {Gnerucci}, A.},
        title = "{A fundamental relation between mass, star formation rate and metallicity in local and high-redshift galaxies}",
      journal = {\mnras},
         year = 2010,
        month = nov,
       volume = {408},
       number = {4},
        pages = {2115-2127},
          doi = {10.1111/j.1365-2966.2010.17291.x},
archivePrefix = {arXiv},
       eprint = {1005.0006},
 primaryClass = {astro-ph.CO},
       adsurl = {https://ui.adsabs.harvard.edu/abs/2010MNRAS.408.2115M}
}

@ARTICLE{Bezanson2015,
       author = {{Bezanson}, Rachel and {Franx}, Marijn and {van Dokkum}, Pieter G.},
        title = "{One Plane for All: Massive Star-forming and Quiescent Galaxies Lie on the Same Mass Fundamental Plane at z \raisebox{-0.5ex}\textasciitilde 0 and z \raisebox{-0.5ex}\textasciitilde 0.7}",
      journal = {\apj},
         year = 2015,
        month = feb,
       volume = {799},
       number = {2},
          eid = {148},
        pages = {148},
          doi = {10.1088/0004-637X/799/2/148},
archivePrefix = {arXiv},
       eprint = {1410.5818},
 primaryClass = {astro-ph.GA},
       adsurl = {https://ui.adsabs.harvard.edu/abs/2015ApJ...799..148B}
}

@ARTICLE{Finlator2008,
       author = {{Finlator}, Kristian and {Dav{\'e}}, Romeel},
        title = "{The origin of the galaxy mass-metallicity relation and implications for galactic outflows}",
      journal = {\mnras},
         year = 2008,
        month = apr,
       volume = {385},
       number = {4},
        pages = {2181-2204},
          doi = {10.1111/j.1365-2966.2008.12991.x},
archivePrefix = {arXiv},
       eprint = {0704.3100},
 primaryClass = {astro-ph},
       adsurl = {https://ui.adsabs.harvard.edu/abs/2008MNRAS.385.2181F}
}

@ARTICLE{Madau2014,
       author = {{Madau}, Piero and {Dickinson}, Mark},
        title = "{Cosmic Star-Formation History}",
      journal = {\araa},
         year = 2014,
        month = aug,
       volume = {52},
        pages = {415-486},
          doi = {10.1146/annurev-astro-081811-125615},
archivePrefix = {arXiv},
       eprint = {1403.0007},
 primaryClass = {astro-ph.CO},
       adsurl = {https://ui.adsabs.harvard.edu/abs/2014ARA&A..52..415M}
}

@ARTICLE{Cullen2019,
       author = {{Cullen}, F. and {McLure}, R.~J. and {Dunlop}, J.~S. and {Khochfar}, S. and {Dav{\'e}}, R. and {Amor{\'\i}n}, R. and {Bolzonella}, M. and {Carnall}, A.~C. and {Castellano}, M. and {Cimatti}, A. and {Cirasuolo}, M. and {Cresci}, G. and {Fynbo}, J.~P.~U. and {Fontanot}, F. and {Gargiulo}, A. and {Garilli}, B. and {Guaita}, L. and {Hathi}, N. and {Hibon}, P. and {Mannucci}, F. and {Marchi}, F. and {McLeod}, D.~J. and {Pentericci}, L. and {Pozzetti}, L. and {Shapley}, A.~E. and {Talia}, M. and {Zamorani}, G.},
        title = "{The VANDELS survey: the stellar metallicities of star-forming galaxies at 2.5 < z < 5.0}",
      journal = {\mnras},
         year = 2019,
        month = aug,
       volume = {487},
       number = {2},
        pages = {2038-2060},
          doi = {10.1093/mnras/stz1402},
archivePrefix = {arXiv},
       eprint = {1903.11081},
 primaryClass = {astro-ph.GA},
       adsurl = {https://ui.adsabs.harvard.edu/abs/2019MNRAS.487.2038C}
}

@ARTICLE{Cullen2021,
       author = {{Cullen}, F. and {Shapley}, A.~E. and {McLure}, R.~J. and {Dunlop}, J.~S. and {Sanders}, R.~L. and {Topping}, M.~W. and {Reddy}, N.~A. and {Amor{\'\i}n}, R. and {Begley}, R. and {Bolzonella}, M. and {Calabr{\`o}}, A. and {Carnall}, A.~C. and {Castellano}, M. and {Cimatti}, A. and {Cirasuolo}, M. and {Cresci}, G. and {Fontana}, A. and {Fontanot}, F. and {Garilli}, B. and {Guaita}, L. and {Hamadouche}, M. and {Hathi}, N.~P. and {Mannucci}, F. and {McLeod}, D.~J. and {Pentericci}, L. and {Saxena}, A. and {Talia}, M. and {Zamorani}, G.},
        title = "{The NIRVANDELS Survey: a robust detection of {\ensuremath{\alpha}}-enhancement in star-forming galaxies at z ≃ 3.4}",
      journal = {\mnras},
         year = 2021,
        month = jul,
       volume = {505},
       number = {1},
        pages = {903-920},
          doi = {10.1093/mnras/stab1340},
archivePrefix = {arXiv},
       eprint = {2103.06300},
 primaryClass = {astro-ph.GA},
       adsurl = {https://ui.adsabs.harvard.edu/abs/2021MNRAS.505..903C}
}

@ARTICLE{Stanton2024,
       author = {{Stanton}, T.~M. and {Cullen}, F. and {McLure}, R.~J. and {Shapley}, A.~E. and {Arellano-C{\'o}rdova}, K.~Z. and {Begley}, R. and {Amor{\'\i}n}, R. and {Barrufet}, L. and {Calabr{\`o}}, A. and {Carnall}, A.~C. and {Cirasuolo}, M. and {Dunlop}, J.~S. and {Donnan}, C.~T. and {Hamadouche}, M.~L. and {Liu}, F.~Y. and {McLeod}, D.~J. and {Pentericci}, L. and {Pozzetti}, L. and {Sanders}, R.~L. and {Scholte}, D. and {Topping}, M.~W.},
        title = "{The NIRVANDELS survey: the stellar and gas-phase mass-metallicity relations of star-forming galaxies at z = 3.5}",
      journal = {\mnras},
         year = 2024,
        month = aug,
       volume = {532},
       number = {3},
        pages = {3102-3119},
          doi = {10.1093/mnras/stae1705},
archivePrefix = {arXiv},
       eprint = {2405.00774},
 primaryClass = {astro-ph.GA},
       adsurl = {https://ui.adsabs.harvard.edu/abs/2024MNRAS.532.3102S}
}

@ARTICLE{Raptis2025,
       author = {{Raptis}, Menelaos and {Trainor}, Ryan F. and {Strom}, Allison L. and {Rudie}, Gwen C. and {Rogers}, Noah S.~J. and {Steidel}, Charles C. and {Maseda}, Michael V. and {von Raesfeld}, Caroline and {Korhonen Cuestas}, Nathalie A.},
        title = "{CECILIA: Ultra-Deep Rest-Optical Spectra of Faint Galaxies at Cosmic Noon}",
      journal = {arXiv e-prints},
         year = 2025,
        month = jul,
          eid = {arXiv:2507.22237},
        pages = {arXiv:2507.22237},
          doi = {10.48550/arXiv.2507.22237},
archivePrefix = {arXiv},
       eprint = {2507.22237},
 primaryClass = {astro-ph.GA},
       adsurl = {https://ui.adsabs.harvard.edu/abs/2025arXiv250722237R}
}

@article{Foreman-Mackey2013,
  title = {emcee: The MCMC Hammer},
  author = {Foreman-Mackey, Daniel and Hogg, David W. and Lang, Dustin and Goodman, Jonathan},
  journal = {Publications of the Astronomical Society of the Pacific},
  volume = {125},
  number = {925},
  pages = {306--312},
  year = {2013},
  doi = {10.1086/670067},
  eprint = {1202.3665},
  archivePrefix = {arXiv},
  primaryClass = {astro-ph.IM},
}

@ARTICLE{Revalski2024,
       author = {{Revalski}, Mitchell and {Rafelski}, Marc and {Henry}, Alaina and {Fossati}, Matteo and {Fumagalli}, Michele and {Dutta}, Rajeshwari and {Pirzkal}, Norbert and {Beckett}, Alexander and {Arrigoni Battaia}, Fabrizio and {Dayal}, Pratika and {D'Odorico}, Valentina and {Lusso}, Elisabeta and {Nedkova}, Kalina V. and {Prichard}, Laura J. and {Papovich}, Casey and {Peroux}, Celine},
        title = "{The MUSE Ultra Deep Field (MUDF). V. Characterizing the Mass{\textendash}Metallicity Relation for Low-mass Galaxies at z {\ensuremath{\sim}} 1{\textendash}2}",
      journal = {\apj},
         year = 2024,
        month = may,
       volume = {966},
       number = {2},
          eid = {228},
        pages = {228},
          doi = {10.3847/1538-4357/ad382c},
archivePrefix = {arXiv},
       eprint = {2403.17047},
 primaryClass = {astro-ph.GA},
       adsurl = {https://ui.adsabs.harvard.edu/abs/2024ApJ...966..228R}
}

@ARTICLE{Ly2016,
       author = {{Ly}, Chun and {Malkan}, Matthew A. and {Rigby}, Jane R. and {Nagao}, Tohru},
        title = "{The Metal Abundances across Cosmic Time (MACT) Survey. II. Evolution of the Mass-metallicity Relation over 8 Billion Years, Using [OIII]4363AA-based Metallicities}",
      journal = {\apj},
         year = 2016,
        month = sep,
       volume = {828},
       number = {2},
          eid = {67},
        pages = {67},
          doi = {10.3847/0004-637X/828/2/67},
archivePrefix = {arXiv},
       eprint = {1602.01098},
 primaryClass = {astro-ph.GA},
       adsurl = {https://ui.adsabs.harvard.edu/abs/2016ApJ...828...67L}
}

@ARTICLE{Gburek2023,
       author = {{Gburek}, Timothy and {Siana}, Brian and {Alavi}, Anahita and {Emami}, Najmeh and {Richard}, Johan and {Freeman}, William R. and {Stark}, Daniel P. and {Snapp-Kolas}, Christopher},
        title = "{The Direct-method Oxygen Abundance of Typical Dwarf Galaxies at Cosmic High Noon}",
      journal = {\apj},
         year = 2023,
        month = may,
       volume = {948},
       number = {2},
          eid = {108},
        pages = {108},
          doi = {10.3847/1538-4357/acb153},
archivePrefix = {arXiv},
       eprint = {2208.05976},
 primaryClass = {astro-ph.GA},
       adsurl = {https://ui.adsabs.harvard.edu/abs/2023ApJ...948..108G}
}

@ARTICLE{bertin1996,
       author = {{Bertin}, E. and {Arnouts}, S.},
        title = "{SExtractor: Software for source extraction.}",
      journal = {\aaps},
         year = 1996,
        month = jun,
       volume = {117},
        pages = {393-404},
          doi = {10.1051/aas:1996164},
       adsurl = {https://ui.adsabs.harvard.edu/abs/1996A&AS..117..393B}
}

@ARTICLE{Clarke2024,
       author = {{Clarke}, Leonardo and {Shapley}, Alice E. and {Sanders}, Ryan L. and {Topping}, Michael W. and {Brammer}, Gabriel B. and {Bento}, Trinity and {Reddy}, Naveen A. and {Kehoe}, Emily},
        title = "{The Star-forming Main Sequence in JADES and CEERS at z > 1.4: Investigating the Burstiness of Star Formation}",
      journal = {\apj},
         year = 2024,
        month = dec,
       volume = {977},
       number = {1},
          eid = {133},
        pages = {133},
          doi = {10.3847/1538-4357/ad8ba4},
archivePrefix = {arXiv},
       eprint = {2406.05178},
 primaryClass = {astro-ph.GA},
       adsurl = {https://ui.adsabs.harvard.edu/abs/2024ApJ...977..133C}
}

@ARTICLE{Trainor2025,
       author = {{Trainor}, Ryan F. and {Lamb}, Noah R. and {Steidel}, Charles C. and {Chen}, Yuguang and {Erb}, Dawn K. and {Trenholm}, Elizabeth and {McClain}, Rebecca L. and {Kovach}, Io},
        title = "{The Lyman-alpha Halos of Galaxies at z=2-3 in the Keck Baryonic Structure Survey}",
      journal = {arXiv e-prints},
         year = 2025,
        month = may,
          eid = {arXiv:2505.15881},
        pages = {arXiv:2505.15881},
          doi = {10.48550/arXiv.2505.15881},
archivePrefix = {arXiv},
       eprint = {2505.15881},
 primaryClass = {astro-ph.GA},
       adsurl = {https://ui.adsabs.harvard.edu/abs/2025arXiv250515881T}
}

@ARTICLE{Carnall2018,
       author = {{Carnall}, A.~C. and {McLure}, R.~J. and {Dunlop}, J.~S. and {Dav{\'e}}, R.},
        title = "{Inferring the star formation histories of massive quiescent galaxies with BAGPIPES: evidence for multiple quenching mechanisms}",
      journal = {\mnras},
         year = 2018,
        month = nov,
       volume = {480},
       number = {4},
        pages = {4379-4401},
          doi = {10.1093/mnras/sty2169},
archivePrefix = {arXiv},
       eprint = {1712.04452},
 primaryClass = {astro-ph.GA},
       adsurl = {https://ui.adsabs.harvard.edu/abs/2018MNRAS.480.4379C}
}

@ARTICLE{Kroupa2001,
       author = {{Kroupa}, Pavel},
        title = "{On the variation of the initial mass function}",
      journal = {\mnras},
         year = 2001,
        month = apr,
       volume = {322},
       number = {2},
        pages = {231-246},
          doi = {10.1046/j.1365-8711.2001.04022.x},
archivePrefix = {arXiv},
       eprint = {astro-ph/0009005},
 primaryClass = {astro-ph},
       adsurl = {https://ui.adsabs.harvard.edu/abs/2001MNRAS.322..231K}
}

@ARTICLE{Gordon2003,
       author = {{Gordon}, Karl D. and {Clayton}, Geoffrey C. and {Misselt}, K.~A. and {Landolt}, Arlo U. and {Wolff}, Michael J.},
        title = "{A Quantitative Comparison of the Small Magellanic Cloud, Large Magellanic Cloud, and Milky Way Ultraviolet to Near-Infrared Extinction Curves}",
      journal = {\apj},
         year = 2003,
        month = sep,
       volume = {594},
       number = {1},
        pages = {279-293},
          doi = {10.1086/376774},
archivePrefix = {arXiv},
       eprint = {astro-ph/0305257},
 primaryClass = {astro-ph},
       adsurl = {https://ui.adsabs.harvard.edu/abs/2003ApJ...594..279G}
}

@ARTICLE{Sanders2025,
       author = {{Sanders}, Ryan L. and {Shapley}, Alice E. and {Topping}, Michael W. and {Reddy}, Naveen A. and {Berg}, Danielle A. and {Khostovan}, Ali Ahmad and {Bouwens}, Rychard J. and {Brammer}, Gabriel and {Carnall}, Adam C. and {Cullen}, Fergus and {Dav{\'e}}, Romeel and {Dunlop}, James S. and {Ellis}, Richard S. and {F{\"o}rster Schreiber}, N.~M. and {Furlanetto}, Steven R. and {Glazebrook}, Karl and {Illingworth}, Garth D. and {Jones}, Tucker and {Kriek}, Mariska and {McLeod}, Derek J. and {McLure}, Ross J. and {Narayanan}, Desika and {Oesch}, Pascal A. and {Pahl}, Anthony J. and {Pettini}, Max and {Schaerer}, Daniel and {Stark}, Daniel P. and {Steidel}, Charles C. and {Tang}, Mengtao and {Clarke}, Leonardo and {Donnan}, Callum T. and {Kehoe}, Emily},
        title = "{The AURORA Survey: High-Redshift Empirical Metallicity Calibrations from Electron Temperature Measurements at z=2-10}",
      journal = {arXiv e-prints},
         year = 2025,
        month = aug,
          eid = {arXiv:2508.10099},
        pages = {arXiv:2508.10099},
          doi = {10.48550/arXiv.2508.10099},
archivePrefix = {arXiv},
       eprint = {2508.10099},
 primaryClass = {astro-ph.GA},
       adsurl = {https://ui.adsabs.harvard.edu/abs/2025arXiv250810099S}
}

@ARTICLE{Rogers2025,
       author = {{Rogers}, Noah S.~J. and {Strom}, Allison L. and {Rudie}, Gwen C. and {Trainor}, Ryan F. and {von Raesfeld}, Caroline and {Raptis}, Menelaos and {Korhonen Cuestas}, Nathalie A. and {Miller}, Tim B. and {Steidel}, Charles C. and {Maseda}, Michael V. and {Chen}, Yuguang and {Law}, David R.},
        title = "{CECILIA: Gas-Phase Physical Conditions and Multi-Element Chemistry at Cosmic Noon}",
      journal = {arXiv e-prints},
         year = 2025,
        month = sep,
          eid = {arXiv:2509.18257},
        pages = {arXiv:2509.18257},
archivePrefix = {arXiv},
       eprint = {2509.18257},
 primaryClass = {astro-ph.GA},
       adsurl = {https://ui.adsabs.harvard.edu/abs/2025arXiv250918257R}
}

@ARTICLE{Jakobsen2022,
       author = {{Jakobsen}, P. and {Ferruit}, P. and {Alves de Oliveira}, C. and {Arribas}, S. and {Bagnasco}, G. and {Barho}, R. and {Beck}, T.~L. and {Birkmann}, S. and {B{\"o}ker}, T. and {Bunker}, A.~J. and {Charlot}, S. and {de Jong}, P. and {de Marchi}, G. and {Ehrenwinkler}, R. and {Falcolini}, M. and {Fels}, R. and {Franx}, M. and {Franz}, D. and {Funke}, M. and {Giardino}, G. and {Gnata}, X. and {Holota}, W. and {Honnen}, K. and {Jensen}, P.~L. and {Jentsch}, M. and {Johnson}, T. and {Jollet}, D. and {Karl}, H. and {Kling}, G. and {K{\"o}hler}, J. and {Kolm}, M. -G. and {Kumari}, N. and {Lander}, M.~E. and {Lemke}, R. and {L{\'o}pez-Caniego}, M. and {L{\"u}tzgendorf}, N. and {Maiolino}, R. and {Manjavacas}, E. and {Marston}, A. and {Maschmann}, M. and {Maurer}, R. and {Messerschmidt}, B. and {Moseley}, S.~H. and {Mosner}, P. and {Mott}, D.~B. and {Muzerolle}, J. and {Pirzkal}, N. and {Pittet}, J. -F. and {Plitzke}, A. and {Posselt}, W. and {Rapp}, B. and {Rauscher}, B.~J. and {Rawle}, T. and {Rix}, H. -W. and {R{\"o}del}, A. and {Rumler}, P. and {Sabbi}, E. and {Salvignol}, J. -C. and {Schmid}, T. and {Sirianni}, M. and {Smith}, C. and {Strada}, P. and {te Plate}, M. and {Valenti}, J. and {Wettemann}, T. and {Wiehe}, T. and {Wiesmayer}, M. and {Willott}, C.~J. and {Wright}, R. and {Zeidler}, P. and {Zincke}, C.},
        title = "{The Near-Infrared Spectrograph (NIRSpec) on the James Webb Space Telescope. I. Overview of the instrument and its capabilities}",
      journal = {\aap},
         year = 2022,
        month = may,
       volume = {661},
          eid = {A80},
        pages = {A80},
          doi = {10.1051/0004-6361/202142663},
archivePrefix = {arXiv},
       eprint = {2202.03305},
 primaryClass = {astro-ph.IM},
       adsurl = {https://ui.adsabs.harvard.edu/abs/2022A&A...661A..80J}
}

@misc{JDox,
  author       = {{STScI}},
  title        = {JWST User Documentation},
  howpublished = {\url{https://jwst-docs.stsci.edu/}},
  note         = {Baltimore, MD: Space Telescope Science Institute},
  year         = {2024},
}

@ARTICLE{Theios2019,
       author = {{Theios}, Rachel L. and {Steidel}, Charles C. and {Strom}, Allison L. and {Rudie}, Gwen C. and {Trainor}, Ryan F. and {Reddy}, Naveen A.},
        title = "{Dust Attenuation, Star Formation, and Metallicity in z {\ensuremath{\sim}} 2-3 Galaxies from KBSS-MOSFIRE}",
      journal = {\apj},
         year = 2019,
        month = jan,
       volume = {871},
       number = {1},
          eid = {128},
        pages = {128},
          doi = {10.3847/1538-4357/aaf386},
archivePrefix = {arXiv},
       eprint = {1805.00016},
 primaryClass = {astro-ph.GA},
       adsurl = {https://ui.adsabs.harvard.edu/abs/2019ApJ...871..128T}
}

@ARTICLE{Juneau2014,
       author = {{Juneau}, St{\'e}phanie and {Bournaud}, Fr{\'e}d{\'e}ric and {Charlot}, St{\'e}phane and {Daddi}, Emanuele and {Elbaz}, David and {Trump}, Jonathan R. and {Brinchmann}, Jarle and {Dickinson}, Mark and {Duc}, Pierre-Alain and {Gobat}, Raphael and {Jean-Baptiste}, Ingrid and {Le Floc'h}, {\'E}meric and {Lehnert}, M.~D. and {Pacifici}, Camilla and {Pannella}, Maurilio and {Schreiber}, Corentin},
        title = "{Active Galactic Nuclei Emission Line Diagnostics and the Mass-Metallicity Relation up to Redshift z \raisebox{-0.5ex}\textasciitilde 2: The Impact of Selection Effects and Evolution}",
      journal = {\apj},
         year = 2014,
        month = jun,
       volume = {788},
       number = {1},
          eid = {88},
        pages = {88},
          doi = {10.1088/0004-637X/788/1/88},
archivePrefix = {arXiv},
       eprint = {1403.6832},
 primaryClass = {astro-ph.GA},
       adsurl = {https://ui.adsabs.harvard.edu/abs/2014ApJ...788...88J}
}

@ARTICLE{Coil2015,
       author = {{Coil}, Alison L. and {Aird}, James and {Reddy}, Naveen and {Shapley}, Alice E. and {Kriek}, Mariska and {Siana}, Brian and {Mobasher}, Bahram and {Freeman}, William R. and {Price}, Sedona H. and {Shivaei}, Irene},
        title = "{The MOSDEF Survey: Optical Active Galactic Nucleus Diagnostics at z \raisebox{-0.5ex}\textasciitilde 2.3}",
      journal = {\apj},
         year = 2015,
        month = mar,
       volume = {801},
       number = {1},
          eid = {35},
        pages = {35},
          doi = {10.1088/0004-637X/801/1/35},
archivePrefix = {arXiv},
       eprint = {1409.6522},
 primaryClass = {astro-ph.GA},
       adsurl = {https://ui.adsabs.harvard.edu/abs/2015ApJ...801...35C}
}

@ARTICLE{deGraaff2024,
       author = {{de Graaff}, Anna and {Rix}, Hans-Walter and {Carniani}, Stefano and {Suess}, Katherine A. and {Charlot}, St{\'e}phane and {Curtis-Lake}, Emma and {Arribas}, Santiago and {Baker}, William M. and {Boyett}, Kristan and {Bunker}, Andrew J. and {Cameron}, Alex J. and {Chevallard}, Jacopo and {Curti}, Mirko and {Eisenstein}, Daniel J. and {Franx}, Marijn and {Hainline}, Kevin and {Hausen}, Ryan and {Ji}, Zhiyuan and {Johnson}, Benjamin D. and {Jones}, Gareth C. and {Maiolino}, Roberto and {Maseda}, Michael V. and {Nelson}, Erica and {Parlanti}, Eleonora and {Rawle}, Tim and {Robertson}, Brant and {Tacchella}, Sandro and {{\"U}bler}, Hannah and {Williams}, Christina C. and {Willmer}, Christopher N.~A. and {Willott}, Chris},
        title = "{Ionised gas kinematics and dynamical masses of z {\ensuremath{\gtrsim}} 6 galaxies from JADES/NIRSpec high-resolution spectroscopy}",
      journal = {\aap},
         year = 2024,
        month = apr,
       volume = {684},
          eid = {A87},
        pages = {A87},
          doi = {10.1051/0004-6361/202347755},
archivePrefix = {arXiv},
       eprint = {2308.09742},
 primaryClass = {astro-ph.GA},
       adsurl = {https://ui.adsabs.harvard.edu/abs/2024A&A...684A..87D}
}

@ARTICLE{Marszewski2025,
       author = {{Marszewski}, Andrew and {Faucher-Gigu{\`e}re}, Claude-Andr{\'e} and {Feldmann}, Robert and {Sun}, Guochao},
        title = "{Explaining the Weak Evolution of the High-redshift Mass{\textendash}Metallicity Relation with Galaxy Burst Cycles}",
      journal = {\apjl},
         year = 2025,
        month = sep,
       volume = {991},
       number = {1},
          eid = {L4},
        pages = {L4},
          doi = {10.3847/2041-8213/adf74b},
archivePrefix = {arXiv},
       eprint = {2505.22712},
 primaryClass = {astro-ph.GA},
       adsurl = {https://ui.adsabs.harvard.edu/abs/2025ApJ...991L...4M}
}

\end{document}